\documentclass[pre,groupedaddress,showpacs]{revtex4}
\usepackage[]{graphicx} 
\usepackage{latexsym,amsmath,amsfonts,amssymb,bm,empheq}  
\usepackage{multirow}
\usepackage{threeparttable}
\usepackage{here}
\usepackage{booktabs}

\begin{document}
\title{Replica Field Theory for a Generalized Franz--Parisi Potential of Inhomogeneous Glassy Systems: New Closure and the Associated Self-Consistent Equation}

\author{Hiroshi Frusawa}
\email{frusawa.hiroshi@kochi-tech.ac.jp}
\affiliation{Laboratory of Statistical Physics, Kochi University of Technology, Tosa-Yamada, Kochi 782-8502, Japan.}
\date{\today}
\begin{abstract}
On approaching the dynamical transition temperature, supercooled liquids show heterogeneity over space and time. Static replica theory investigates the dynamical crossover in terms of the free energy landscape (FEL). Two kinds of static approaches have provided a self-consistent equation for determining this crossover, similar to the mode coupling theory for glassy dynamics. One uses the Morita--Hiroike formalism of the liquid state theory, whereas the other relies on the density functional theory (DFT). Each of the two approaches has advantages in terms of perturbative field theory. Here, we develop a replica field theory that has the benefits from both formulations. We introduce the generalized Franz--Parisi potential to formulate a correlation functional. Considering fluctuations around an inhomogeneous density determined by the Ramakrishnan--Yussouf DFT, we find a new closure as the stability condition of the correlation functional. The closure leads to the self-consistent equation involving the triplet direct correlation function. The present field theory further helps us study the FEL beyond the mean-field approximation.
\end{abstract}

\maketitle
\begin{figure}[H]
\begin{center}
\includegraphics[width=7cm]{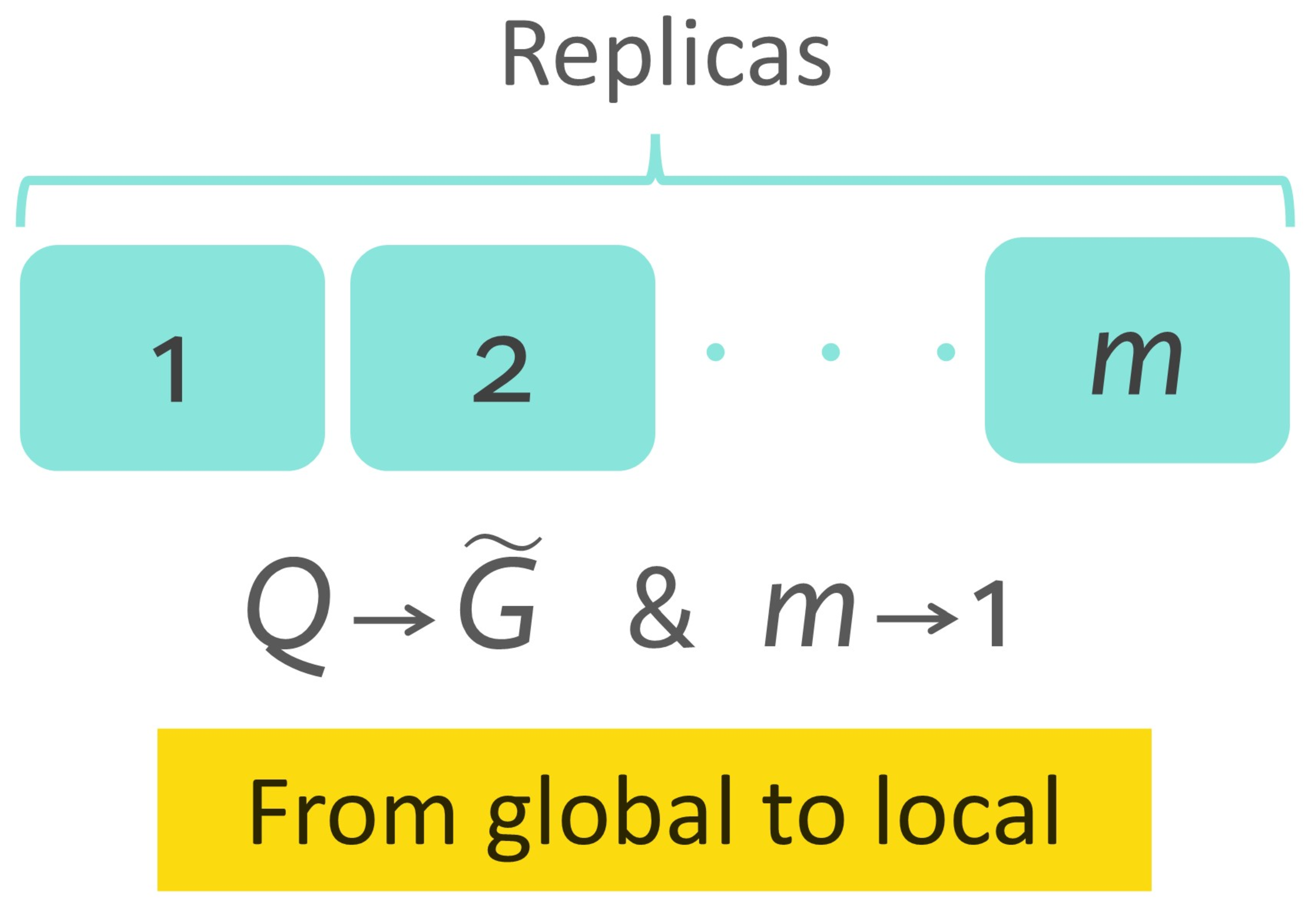}
\end{center}
\end{figure}


\section{Introduction}

Glass, an~amorphous solid with elasticity, has a microscopic structure in which localized particles oscillate around their mean positions of a random lattice~\cite{g1,g2,g3,g4,g5,g6}.
The spatial randomness is self-generated by the particle localization that breaks translational symmetry.
A remarkable feature of the random structure is that glass microscopically lacks the long-range order and is similar to liquid in terms of density--density correlations~\cite{g1,g2,g3,g4,g5,g6}.
As a precursor to the random structure of glass, supercooled liquids show heterogeneity over space and time~\cite{g1,g2,g3,g4,g5,g6,het1,het2,het3,het4,het5}.
The dynamical heterogeneity and facilitation~\cite{g4,het1,het2,het3,het4,het5,fac1,fac2,fac4} emerges on approaching the dynamical transition temperature ($T_d$), accompanied by the crossover from relaxational to activated dynamics.
For $T>T_d$, transport is not collective on a large scale.
For $T\rightarrow T_d$, the~system gets stuck in a glassy metastable state, and~the dynamical behaviors of supercooled liquids exhibit features such as a two-step decay with a first relaxation ($\beta$-relaxation) to a plateau followed by a stretched exponential relaxation ($\alpha$-relaxation) of density fluctuations.
Along with this dynamical crossover, the~dynamics become progressively heterogeneous and correlated in space.
For $T<T_d$, the~relaxation times of the two-step decay increase rapidly despite slight changes in the disordered~microstructure.

Various theories have tried to explain the dynamical heterogeneity and facilitation, as well as the dynamical crossover at $T_d$.
These include either elasticity theory or kinetically constrained models focusing on the dynamical facilitation~\cite{g4,fac1,fac2,fac4}, and~the mode coupling theory (MCT) \cite{mct1,mct2}, a~dynamical theory relevant when approaching $T_d$ from the liquid phase.
The MCT describes the onset of the two-step relaxation above $T_d$ and predicts the divergence of $\beta$-relaxation time at $T_d$.
Extension to the inhomogeneous MCT further allows us to describe a growing dynamical heterogeneity using a time-dependent three-or four-point correlation function~\cite{g2,het1,het2,het3,het4,het5}.
Yet the dynamical transition temperature $T_d$ is higher than the glass transition temperature observed in simulation and experimental studies.
An interpretation of this discrepancy is that a mean-field description of the MCT is beyond the scope of the barrier-dominated dynamics between metastable states, though~applicable to the relaxation dynamics within a metastable state.
The divergent behavior of the two-step decay predicted by the MCT at $T_d$ becomes incomplete because of the activated events remaining in actual liquids for $T\leq T_d$ \cite{mct1,mct2}.

The activation dynamics dominant below $T_d$ are due to transitions between metastable states~\cite{mct1,mct2,rfot1,rfot2,rfot3,rfot4}.
Therefore, the~dynamical crossover implies the emergence of many metastable states at $T_d$ or the appearance of a free energy landscape (FEL) characterized by an exponentially large number of metastable states below $T_d$.
From the thermodynamic point of view, we can describe the characteristic of the FEL using the configurational entropy obtained from the logarithm of the number of metastable states~\cite{g1,g2,rfot1,rfot2,rfot3,rfot4}.
The Adam--Gibbs relation provides results in quantitative agreement with simulation and experimental results, relating the drastic changes in the relaxation time and viscosity to the decrease of the configurational entropy on approaching the glass transition temperature~\cite{rfot1,rfot2,rfot3,rfot4,adam1,adam2}.
For example, simulation studies on mixtures interacting via the Lennard-Jones potential and its repulsive counterpart, the~WCA one, demonstrate that these systems exhibit quite different dynamics despite having nearly identical structures~\cite{adam3,adam4,adam5,adam6,adam7,adam8}.
Such a large difference in the dynamics is ascribable to a considerable gap between the configurational entropies while making a slight difference between the two-point correlation functions.
Previous investigations confirmed that the configurational entropies associated with correlation functions differ greatly between the Lennard-Jones and WCA mixtures despite the structural similarity, therefore predicting the distinct dynamical behaviors from the Adam--Gibbs relation~\cite{adam3,adam4,adam5,adam6,adam7,adam8}.

Static approaches, other than dynamical ones such as the MCT, are relevant to investigate the FEL or the configurational entropy~\cite{rfot1,rfot2,rfot3,rfot4,semi1,semi2,semi3,scages,mor1,mor2,mor3,mor4,mor5,mor6,dfg1,dfg2,dfg3,dfg4,dfg5,dfg6,dfg7,dfg8,dfg9,dfg10,dfg11,dfg12,dfg13,dfg14,dfr1,dfr2,dfr3,dfr4,dfr5}.
These include replica theory~\cite{rfot1,rfot2,rfot3,rfot4,semi1,semi2,semi3,scages,mor1,mor2,mor3,mor4,mor5,mor6}, density functional theory (DFT) \cite{dfg1,dfg2,dfg3,dfg4,dfg5,dfg6,dfg7,dfg8,dfg9,dfg10,dfg11,dfg12,dfg13,dfg14}, and~a combination of the replica theory and DFT~\cite{dfr1,dfr2,dfr3,dfr4,dfr5}.
The static theories commonly focus on local minima of free-energy functionals without considering fluctuations due to the mean-field approximation.
On the one hand, the~DFT determines the metastable state by exploring a local minimum of the free-energy density functional~\cite{dfg1,dfg2,dfg3,dfg4,dfg5,dfg6,dfg7,dfg8,dfg9,dfg10,dfg11,dfg12,dfg13,dfg14}.
Given the inhomogeneous density distribution as overlapping Gaussians centered around a random lattice, previous studies have confirmed that Gaussian distribution with a large spread creates the optimum density profile.
The low degree of localization around the random lattice is consistent with experimental and simulation results.
On the other hand, replica theory considers a system of coupled $m$-replicas of the original system~\cite{rfot1,rfot2,rfot3,rfot4,semi1,semi2,semi3,scages,mor1,mor2,mor3,mor4,mor5,mor6,dfr1,dfr2,dfr3,dfr4,dfr5}.
The replica free-energy functional depends on a two-point correlation function between two copies (an inter-replica correlation function), an~order parameter measuring the degree of similarity between two typical configurations.
We obtain the correct result by taking the limit of $m\rightarrow 1$ with the inter-replica coupling switched off.
While the order parameter goes to zero in the liquid state without the inter-replica coupling, the~order parameter in an ergodicity-broken phase has a finite value because two copies remain highly correlated even after switching off the inter-replica coupling.
The replica theory has successfully explained experimental and simulation results using the following four approximations: the small cage expansion~\cite{rfot1,rfot2,rfot3,rfot4,semi3,scages}, the~effective potential approximation~\cite{rfot1,rfot2,rfot3,rfot4,semi3,scages}, the~replicated hypernetted-chain (RHNC) approximation~\cite{mor1,mor2,mor3,mor4,mor5,mor6}, and~the third-order functional expansion in DFT~\cite{dfr4,dfr5,dft1,dft2,dft3,ry}.
While the first two are perturbation methods with the local cage size as a reference scale, the~last two approximations cover those of the liquid-state theory~\cite{ls1,ls2,morita}.

The Franz--Parisi (FP) potential obtained in the RHNC approximation serves as a starting point for this paper.
The FP potential~\cite{fp1,fp2,fp3,fp4,fp5,fp6,fp7,fp8,fp9,fp10,fp11} is a function of overlap $Q$, a~weighted average over the system of the two-point correlation function, and~plays the same role as the Landau free energy of a global parameter $Q$ that indicates a distance between the two copies in configuration space.
Theoretical and simulation studies have demonstrated that the FP potential reproduces the temperature evolution of FELs, just like the Landau free energy~\cite{fp1,fp2,fp3,fp4,fp5,fp6,fp7,fp8,fp9,fp10,fp11}.
With decreasing temperature, the~FP potential develops a secondary minimum for $Q>0$ representing a metastable state.
Considering $Q=0$ in the liquid state, we can see that the potential difference, $V(Q)-V(0)$, corresponds to the entropic cost of localizing the system in a single metastable state (i.e., the~configurational entropy).

In this paper, we generalize the FP potential by fixing an inter-replica correlation function instead of the overlap $Q$.
We formulate the generalized FP potential by developing a new framework that is beneficial to investigate the FEL while considering inhomogeneous supercooled liquids with the help of field theoretical method.
A field theory combining the DFT~\cite{dfr1,dfr2,dfr3,dfr4,dfr5,dft1,dft2,dft3,ry} and replica theory~\cite{rfot1,rfot2,rfot3,rfot4,semi1,semi2,semi3,scages,mor1,mor2,mor3,mor4,mor5,mor6,dfr1,dfr2,dfr3,dfr4,dfr5} forms the basis of our framework.
There are two requirements to be satisfied by the field theory and the associated functional.
The first requirement is that the developed framework can consider inhomogeneous systems.
The second requirement is that the generalized FP functional applies to non-equilibrium states away from metastable states.
To meet the requirements, this paper presents the correlation functional theory that provides the generalized FP potential functional without going through the Morita-Hiroike functional~\cite{mor1,mor2,mor3,mor4,mor5,mor6,ls1,ls2,morita}.
The generalized FP potential has three features as a functional of density and correlation function.
First, this potential is a functional of metastable density that becomes equal to that of the DFT in the limit of $m\rightarrow 1$.
Second, the~field-theoretical perturbation method allows us to have a new correlation functional different from the Morita--Hiroike one while maintaining consistency with the liquid theory in that the approximate form reduces to the RHNC functional.
Last, the~potential functional of a given inter-replica correlation function has a minimum where a new closure reducible to the RHNC approximation~\cite{mor1,mor2,mor3,mor4,mor5,mor6,ls1,ls2} holds.
A remarkable result is that an approximation of the new closure yields the self-consistent equation for a non-ergodicity parameter that includes the triplet direct correlation function (DCF)~\cite{ls1,ls2,dcf1,dcf2,dcf3}, similar to that formulated by either the MCT~\cite{mct1,mct2} or the replica theory~\cite{mor5,dfr4,dfr5}, respectively.

The paper is organized as follows.
In Section~\ref{section fp}, we define the generalized FP potential.
Comparison between the generalized and original FP potentials clarifies what we modify through the generalization.
Section~\ref{section main} summarizes the theoretical results consisting of four parts as follows: relation for obtaining the generalized FP potential from the grand potential of $m$-replica system with inter-replica correlation function fixed ({\itshape Result 1}
); functional form of the constrained grand potential ({\itshape Result 2}); new closure for two-point correlation function ({\itshape Result 3}); the associated self-consistent equation for a non-ergodicity parameter ({\itshape Result 4}).
We obtain the generalized FP potential from {\itshape Result 2} with the help of the relation in {\itshape Result 1}.
The extremum condition of this potential yields a new closure in {\itshape Result 3}.
It also turns out that a self-consistent equation obtained from an approximate form of the closure involves the triplet DCF as presented in {\itshape Result 4}.
In Section~\ref{functional integrals}, we calculate the perturbative terms using a strong-coupling perturbation theory developed for obtaining {\itshape Result 2} (see Appendix \ref{sc}).
In the saddle-point approximation, the~strong-coupling perturbation theory provides the correlation functional form of the constrained grand potential given in {\itshape Result 2}.
In Section~\ref{section con}, we make some concluding~remarks.

\section{Generalized Franz--Parisi (FP) Potential}\label{section fp}
We generalize the FP potential in comparison with its original definition.
\subsection{The Original FP~Potential}
Let $\mathcal{C}_a$ be a configuration that represents a set of $N$-particle positions
, $\{\bm{r}_{a,i}\}_{i=1,\cdots,N}$, in~replica $a$ ($1\leq a\leq m$) when considering $m$ copies of the liquid.
The overlap $\widehat{Q}(\widehat{\rho}_a,\widehat{\rho}_b)$ ($a\neq b$) measures the degree of similarity between a pair of replicas using the microscopic density (or the so-called density ``operator''~\cite{text,fred1,fred2,matsen,euclid,f0,f2,f4,woo,russia}) in replica $a$, $\widehat{\rho}_a^{(N)}(\bm{r})=\sum_{i=1}^N\delta(\bm{r}-\bm{r}_{a,i})$.
We define that
\begin{flalign}
\widehat{Q}(\widehat{\rho}_a,\widehat{\rho}_b)
=\frac{1}{N}\int\!\!\!\int d\bm{r}\,d\bm{r}'
\widehat{\rho}_a^{(N)}(\bm{r})\widehat{\rho}_b^{(N)}(\bm{r}')\eta(\bm{r}-\bm{r}'),
\label{def q}
\end{flalign}
where a distribution function $\eta(\bm{r})$ specifies the spatial averaging performed over a finite range; for example, we have $\eta(\bm{r})=\Theta(a-|\bm{r}|)$ using the Heaviside function
$\Theta(r)$ and particle diameter $a$ \cite{fp1,fp2,fp3,fp4,fp5,fp6,fp7,fp8,fp9,fp10,fp11}.

The FP potential $V(Q)$ is obtained in two steps~\cite{fp1,fp2,fp3,fp4,fp5,fp6,fp7,fp8,fp9,fp10,fp11}.
First, we fix a reference configuration $\widehat{\rho}_1$ of replica 1, which plays the role of quenched variable in the effective potential $V(Q,\widehat{\rho}_1)$ as seen from the following definition:
\begin{flalign}
e^{-\beta NV^+(Q,\widehat{\rho}_1)}
&=\sum_{\mathcal{C}_a}\,
e^{-\beta U_a(\widehat{\rho}_1,\widehat{\rho}_a)}
\delta\left[\,Q-\widehat{Q}(\widehat{\rho}_1,\widehat{\rho}_a)
\right],
\label{def vqc1}\\
V(Q,\widehat{\rho}_1)&=\lim_{U_{\mathrm{inter}}\rightarrow 0}V^+(Q,\widehat{\rho}_1),
\label{def vqc2}
\end{flalign}
where $\sum_{\mathcal{C}_a}$ denotes $(1/N!)\int\cdots\int d\bm{r}_{a,1}\cdots d\bm{r}_{a,N}$ in the canonical ensemble of replica $a$ for $a\geq2$, $\beta$ the inverse thermal energy $(k_BT)^{-1}$, and~$U_a(\widehat{\rho}_1,\widehat{\rho}_a)$ the interaction energy of replica $a$ that is the sum of intra-replica interaction energy $U_{\mathrm{intra}}(\widehat{\rho}_a)$ and inter-replica one $U_{\mathrm{inter}}(\widehat{\rho}_1,\widehat{\rho}_a)$:
\begin{equation}
U_a(\widehat{\rho}_1,\widehat{\rho}_a)=U_{\mathrm{intra}}(\widehat{\rho}_a)+U_{\mathrm{inter}}(\widehat{\rho}_1,\widehat{\rho}_a),
\label{sum}
\end{equation}
where
\begin{flalign}
&U_{\mathrm{intra}}(\widehat{\rho}_a)
=\frac{1}{2}
\int\!\!\!\int d\bm{r}d\bm{r}'\left\{
\widehat{\rho}_a^{(N)}(\bm{r})
v(\bm{r}-\bm{r}')\widehat{\rho}_a^{(N)}(\bm{r}')
-\widehat{\rho}_a^{(N)}(\bm{r})v(\bm{r}-\bm{r}')
\delta(\bm{r}-\bm{r}')
\right\},
\label{def u intra}\\
&U_{\mathrm{inter}}(\widehat{\rho}_1,\widehat{\rho}_a)
=\int\!\!\!\int d\bm{r}d\bm{r}'
\widehat{\rho}_1^{(N)}(\bm{r})
\widetilde{v}(\bm{r}-\bm{r}')\widehat{\rho}_a^{(N)}(\bm{r}'),
\label{def u inter}
\end{flalign}
using the intra-replica interaction potential $v(\bm{r})$ and the inter-replica one $\widetilde{v}(\bm{r})$.
It is noted that the effective potential $V(Q,\widehat{\rho}_1)$ is defined in the absence of inter-replica interactions as represented by Equation~(\ref{def vqc2}).

Next, we perform the canonical average of $V(Q,\widehat{\rho}_1)$ over all possible choices for the reference configuration with the statistical weight $p_{\rm{eq}}(\widehat{\rho}_1)$ as follows:
\begin{flalign}
V(Q)&=\sum_{\mathcal{C}_1}\,p_{\rm{eq}}(\widehat{\rho}_1)\,V(Q,\widehat{\rho}_1),
\label{def vq}
\\
p_{\rm{eq}}(\widehat{\rho}_1)&=\frac{e^{-\beta U_{\mathrm{intra}}(\widehat{\rho}_1)}}
{\sum_{\mathcal{C}_1}e^{-\beta U_{\mathrm{intra}}(\widehat{\rho}_1)}}.
\label{def peq}
\end{flalign}
The replica trick allows us to calculate Equation~(\ref{def vq}), thus obtaining the FP potential $V(Q)$ of the Landau~type.

\subsection{Generalization}
Here, we introduce a generalized FP potential $W(\widetilde{G})$ as a functional of prescribed correlation function $\widetilde{G}(\bm{r},\bm{r}')$, instead of the overlap $Q$.
In terms of the Landau theory, we consider a local order parameter, instead of the global one.
We use the grand canonical ensemble represented by the following operator:
\begin{flalign}
\mathrm{Tr}_a\equiv
 \sum_{N=0}^{\infty}\frac{e^{N\beta\mu}}{N!}\,\int d\bm{r}_{a,1}\cdots\int d\bm{r}_{a,N}
=\sum_{N=0}^{\infty}e^{N\beta\mu}\sum_{\mathcal{C}_a},
\label{def tr}
\end{flalign}
where the chemical potential $\beta\mu$ in units of $k_BT$ determines the most probable number $N^*$, thereby providing the uniform density $\overline{\rho}=N^*/V$ common to each replica with volume $V$.

Given a reference configuration $\mathcal{C}_1$ of replica 1, we have the interaction energy $U_a(\widehat{\rho}_1,\widehat{\rho}_a)$ of replica $a$, providing the grand potential $\omega_a(\widehat{\rho}_1)$ of replica $a$ as follows:
\begin{flalign}
e^{-\beta\omega^+_a(\widehat{\rho}_1)}&= \mathrm{Tr}_a\,e^{-\beta U_a(\widehat{\rho}_1,\widehat{\rho}_a)}
\nonumber\\
&=\int D\widetilde{G}\,\mathrm{Tr}_a\,e^{-\beta U_a(\widehat{\rho}_1,\widehat{\rho}_a)}
\prod_{b=1,a}\mathcal{I}_b(\rho,\widehat{\rho})
\,\Delta_a(\widetilde{G},\rho)
\nonumber\\
&=\int D\widetilde{G}\,
e^{-\beta N^*W(\widetilde{G},\widehat{\rho}_1)},
\label{omega a1}\\
\omega_a(\widehat{\rho}_1)
&=\lim_{U_{\mathrm{inter}}\rightarrow 0}\omega^+_a(\widehat{\rho}_1),
\label{omega a2}
\end{flalign}
where the functional integral representation in Equation~(\ref{omega a1}) is obtained from multiplying the right-hand side (rhs) of the first line in Equation~(\ref{omega a1}) by the following identity:
\begin{flalign}
1&=\int D\widetilde{G}\prod_{b=1,a}\int D\rho_b\, 
\prod_{\{\bm{r}\},\{\bm{r}'\}}
\prod_{\{\bm{r}\}}\delta\left[\rho_b(\bm{r})-\widehat{\rho}_b^{(N)}(\bm{r})
\right]\,
\delta\left[\widetilde{G}(\bm{r},\bm{r}')-\rho_1(\bm{r})\rho_a(\bm{r}')
\right]\nonumber\\
&=\int D\widetilde{G}\>
\prod_{b=1,a}\mathcal{I}_b(\rho,\widehat{\rho})
\,\Delta_a(\widetilde{G},\rho).
\label{identity1}
\end{flalign}
Equation~(\ref{identity1}) implies that
\begin{flalign}
&\mathcal{I}_b(\rho,\widehat{\rho})\equiv\int D\rho_b 
\prod_{\{\bm{r}\}}\delta\left[\rho_b(\bm{r})-\widehat{\rho}_b^{(N)}(\bm{r})
\right]=1,
\label{identity2}\\
&\Delta_a(\widetilde{G},\rho)\equiv
\prod_{\{\bm{r}\},\{\bm{r}'\}}
\delta\left[\widetilde{G}(\bm{r},\bm{r}')-\rho_1(\bm{r})\rho_a(\bm{r}')
\right].
\label{identity3}
\end{flalign}
The relation (\ref{identity2}) at $b=1$ represents that only the density distribution $\widehat{\rho}_1^{(N)}(\bm{r})$ is allowed due to a fixed configuration $\mathcal{C}_1$ of replica~1. 

Equations~(\ref{omega a1})--
(\ref{identity3}) reveal that the field-theoretical formulation of the effective potential $W(\widetilde{G},\widehat{\rho}_1)$ can be developed as follows~\cite{text,fred1,fred2,matsen,euclid,f0,f2,f4,woo,russia}:
\begin{flalign}
W(\widetilde{G},\widehat{\rho}_1)
&=\lim_{U_{\mathrm{inter}}\rightarrow 0}W^+(\widetilde{G},\widehat{\rho}_1),
\label{def wgc1}
\end{flalign}
where
\begin{flalign}
e^{-\beta N^*W^+(\widetilde{G},\widehat{\rho}_1)}
&=\mathrm{Tr}_a\,e^{-\beta U_a(\widehat{\rho}_1,\widehat{\rho}_a)}
\prod_{b=1,a}\mathcal{I}_b(\rho,\widehat{\rho})
\,\Delta_a(\widetilde{G},\rho)\nonumber\\
&=\mathcal{I}_1(\rho,\widehat{\rho})\,\mathrm{Tr}_a\,
\mathcal{I}_a(\rho,\widehat{\rho})\,
e^{-\beta U_a(\rho_1,\rho_a)}\,
\Delta_a(\widetilde{G},\rho)
\nonumber\\
&=\mathcal{I}_1(\rho,\widehat{\rho})\int D\rho_a\,
e^{-\beta U_a(\rho_1,\rho_a)}
\,\mathrm{Tr}_a\,\prod_{\{\bm{r}\}}\delta\left[\rho_a(\bm{r})-\widehat{\rho}_a^{(N)}(\bm{r})
\right]\,
\Delta_a(\widetilde{G},\rho)
\nonumber\\
&=\mathcal{I}_1(\rho,\widehat{\rho})
\int D\rho_a\,
e^{-\beta \left\{
U_a(\rho_1,\rho_a)
-T\mathcal{S}_a^{\mathrm{id}}(\rho_a)\right\}}\,
\Delta_a(\widetilde{G},\rho).
\label{def wgc2}
\end{flalign}
In the last line of Equation~(\ref{def wgc2}), we have the ideal gas entropy defined by
\begin{flalign}
-T\mathcal{S}_a^{\mathrm{id}}(\rho_a)=k_BT\int d\bm{r}\,\rho_a(\bm{r})\left\{
\ln\rho_a(\bm{r})-1-\beta\mu
\right\}.
\label{def entropy a}
\end{flalign}
The generalized FP potential $W(\widetilde{G})$ is obtained from the grand canonical average of $W(\widetilde{G},\widehat{\rho}_1)$ for the reference configuration as follows:
\begin{flalign}
\label{def wg}
W(\widetilde{G})&
=\mathrm{Tr}_1\,P_{\rm{eq}}(\widehat{\rho}_1)W(\widetilde{G},\widehat{\rho}_1),\\
\label{results peq}
P_{\rm{eq}}(\widehat{\rho}_1)&=\frac{e^{-\beta U_{\mathrm{intra}}(\widehat{\rho}_1)}}
{\mathrm{Tr}_1\,e^{-\beta U_{\mathrm{intra}}(\widehat{\rho}_1)}},
\end{flalign}
similar to Equations~(\ref{def vq}) and (\ref{def peq}). 
Equation~(\ref{def wg}) clarifies that a given configuration $\widehat{\rho}_1$ plays a role of quenched disorder to another replica $a$ \cite{fp1,fp2,fp3,fp4,fp5,fp6,fp7,fp8,fp9,fp10,fp11}.
Since we consider all possible configurations of $\widehat{\rho}_1$, the~statistical weight $P_{\rm{eq}}(\widehat{\rho}_1)$ is of the Boltzmann form as well as $p_{\rm{eq}}(\widehat{\rho}_1)$ in Equation~(\ref{def peq}).

Several remarks on Equations~(\ref{identity2})--(\ref{results peq}) are in~order:
\begin{itemize}
\item Equation~(\ref{identity3}) tells us that a prescribed correlation field $\widetilde{G}(\bm{r},\bm{r}')$ represents a product $\rho_1(\bm{r})\rho_a(\bm{r}')$ of two instantaneous density distributions in different replicas, or~a statistical realization of density-density correlation function~\cite{ls1,ls2}.
\item To perform the configurational integral $\mathrm{Tr}_a$ in the second line on the rhs of \mbox{Equation~(\ref{def wgc2})}, it is indispensable to introduce the Fourier transform representation of the delta functional using the functional integral over the one-body potential field, which is dual to the density field $\rho_a(\bm{r})$ \cite{f0,f2,f4,woo,russia}.
The ideal gas entropy given by Equation~(\ref{def entropy a}) appears in the last line of Equation~(\ref{def wgc2}) due to the saddle-point approximation of the one-body potential field~\cite{f0,f2,f4,woo,russia}.
\item When different replica particles form complexes because of the attractive inter-replica interactions between them (i.e., $\widetilde{v}(\bm{r})<0$), we have $\widetilde{G}(\bm{r})/\overline{\rho}^2\gg 1$ in an overlapped region (e.g., $|\bm{r}|\leq a$), thereby providing a significant value of overlap $Q$ that is greater than the random overlap obtained from $\widetilde{G}(\bm{r})= \overline{\rho}^2$. The~glassy state preserves an overlapped state due to frozen configurations of particles even after the attractive inter-replica interactions are switched off (i.e., $\widetilde{v}(\bm{r})\rightarrow 0$). The~generalized FP potential $W(\widetilde{G})$ is available to explore such an overlapped state that is locally stable. 
\item It is also noted that the above formalism presented in Equations~(\ref{omega a1})--(\ref{def entropy a}) has been conventionally used for the formulation of continuous field theory~\cite{text,fred1,fred2,matsen,euclid,f0,f2,f4,woo,russia};
the density operator $\widehat{\rho}_b^{(N)}(\bm{r})$ ($b=1,\,a$) has been mapped to a density field $\rho_b(\bm{r})$ using the density functional integral in Equations~(\ref{identity2}) and (\ref{def wgc2}) according to the conventional formalism in statistical field theory~\cite{text} (see also the literature~\cite{fred1,fred2,matsen,euclid,f0,f2,f4,woo,russia,dean1,dean2,dean3} for discussions about the underlying physics of this formal procedure to introduce a continuous density field).
\end{itemize}

\section{Main~Results}\label{section main}
We present four sets of main results based on the strong-coupling perturbation theory (see Appendix \ref{sc} for details).
Figure~\ref{fig1} summarizes the results schematically.

\begin{figure}[H]
\begin{center}
\includegraphics[width=12cm]{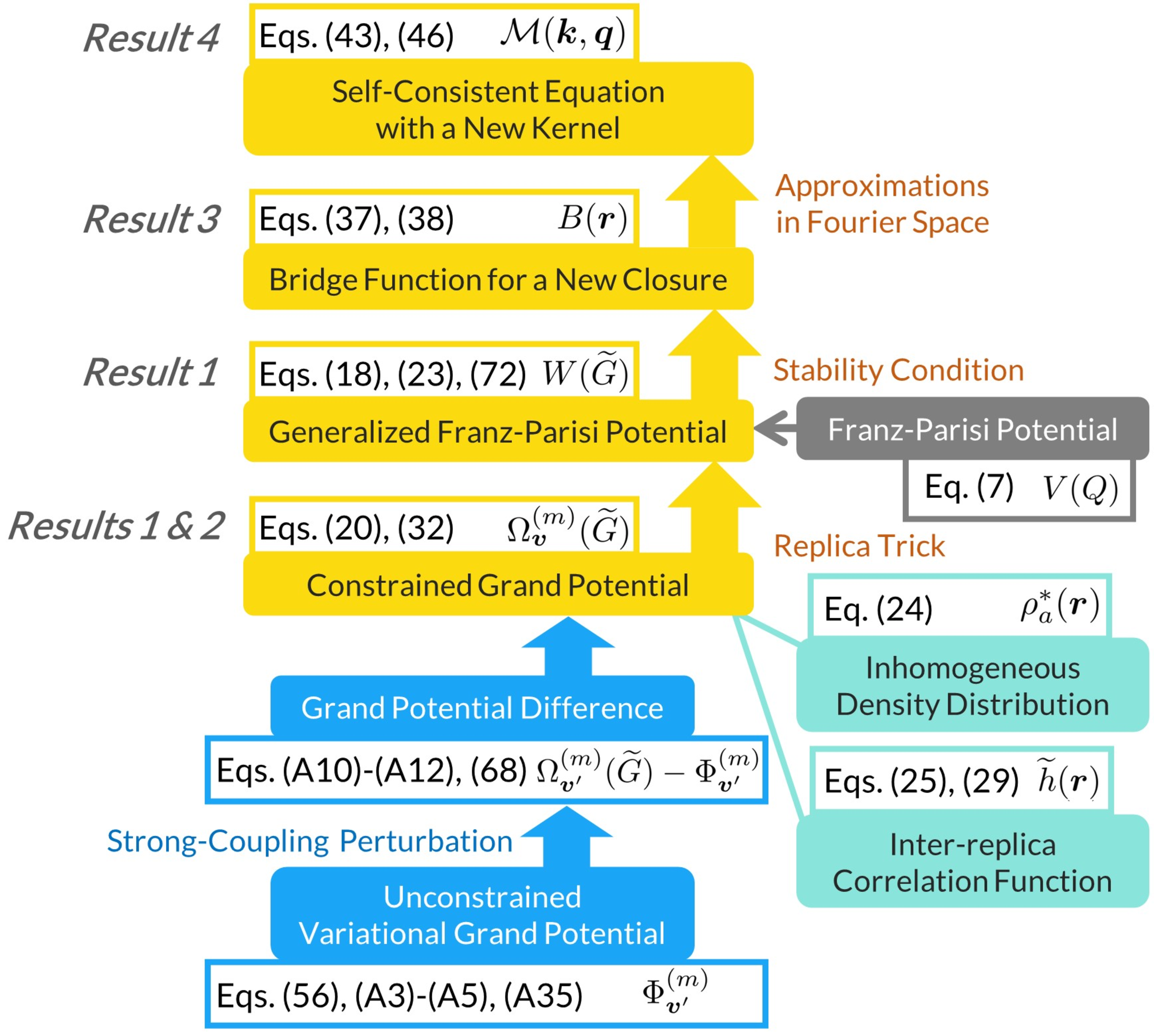}
\end{center}
\caption{A schematic summary of the main results colored orange. In~addition, functional variables are colored green, and~underlying potentials blue or~gray.\label{fig1}}
\end{figure}

\subsection{{\itshape Result~1}: Replica Formalism of the Generalized FP Potential}
Let $\Omega_{\bm{v}}^{(m)}(\widetilde{G})$ be the constrained grand potential  of $m$ replicas defined by
\begin{flalign}
e^{-\beta \Omega_{\bm{v}}^{(m)}(\widetilde{G})}
&=\boldsymbol{\mathrm{Tr}}\,
e^{-\beta U(\bm{v},\boldsymbol{\widehat{\rho}})}
\,\prod_{a=2}^{m}\Delta_a(\widetilde{G},\boldsymbol{\widehat{\rho}})
\nonumber\\
&=\int D\boldsymbol{\rho}\,\boldsymbol{\mathrm{Tr}}\,
e^{-\beta U(\bm{v},\boldsymbol{\rho})}
\prod_{b=1}^m\prod_{\{\bm{r}\}}\delta\left[\rho_b(\bm{r})-\widehat{\rho}_b^{(N)}(\bm{r})
\right]
\,\prod_{a=2}^{m}\Delta_a(\widetilde{G},\rho)
\label{def omega}
\end{flalign}
where $\int D\boldsymbol{\rho}\equiv\prod_{b=1}^m\int D\rho_b$, $\boldsymbol{\mathrm{Tr}}\equiv\prod_{b=1}^m\mathrm{Tr}_b$, the~matrix elements of $\bm{v}$ are $v_{ab}(\bm{r})=0$ ($a\neq b$) and $v_{aa}(\bm{r})=v(\bm{r})$, $\boldsymbol{\widehat{\rho}}=(\widehat{\rho}_1^{(N)},\cdots,\widehat{\rho}_m^{(N)})^{\rm{T}}$, $\boldsymbol{\rho}=(\rho_1,\cdots,\rho_m)^{\rm{T}}$, and~the interaction energy $U(\bm{v},\boldsymbol{\widehat{\rho}})$ in Equation~(\ref{def omega}) is given by
\begin{flalign}
U(\bm{v},\boldsymbol{\widehat{\rho}})
=\frac{1}{2}
\int\!\!\!\int d\bm{r}d\bm{r}'\left\{
\boldsymbol{\widehat{\rho}}(\bm{r})^{\rm{T}}
\bm{v}(\bm{r}-\bm{r}')\boldsymbol{\widehat{\rho}}(\bm{r}')
-\sum_{b=1}^m\widehat{\rho}_b^{(N)}(\bm{r})v(\bm{r}-\bm{r}')
\delta(\bm{r}-\bm{r}')
\right\},
\label{results def u}
\end{flalign}
excluding the intra-replica self-energy.
Incidentally, there are two methods to treat the density functional integral in Equation~(\ref{def omega}) \cite{text,fred1,fred2,matsen,euclid,f0,f2,f4,woo,russia}, both of which will be utilized as seen from Equations~(\ref{potential int}) and (\ref{app dft}).

It is readily seen from Equations~(\ref{def wgc2}) and (\ref{def omega}) that the constrained grand potential $\Omega_{\bm{v}}^{(m)}(\widetilde{G})$ is expressed using $W(\widetilde{G},\widehat{\rho}_1)$ as
\begin{flalign}
e^{-\beta\Omega_{\bm{v}}^{(m)}(\widetilde{G})}=\mathrm{Tr}_1\,e^{-\beta U_{\mathrm{intra}}(\widehat{\rho}_1)-(m-1)\beta N^*W(\widetilde{G},\widehat{\rho}_1)}.
\label{results omega2}
\end{flalign}
The replica trick allows us to have the relation between the constrained grand potential $\Omega_{\bm{v}}^{(m)}$ and the generalized FP potential $W(\widetilde{G})$:
\begin{flalign}
\label{results replica}
N^*W(\widetilde{G})
&=\lim_{m\rightarrow 1}\frac{\partial\Omega_{\bm{v}}^{(m)}(\widetilde{G})}{\partial m},
\end{flalign}
which is the first result ({\itshape Result 1}; see Appendix \ref{appendix replica} for the detailed derivation). 
It is noted that the conventional replica trick proves the necessity of $m\rightarrow 1$ to consider the quenched type of the FP formalism, though~it has been physically motivated to take the limit of $m\rightarrow 1$ based on the Monasson formalism~\cite{rfot2,adam2,fp11}.

\subsection{{\itshape Result~2}: The Constrained Grand Potential Functional of {\itshape m} Replicas in an Inhomogeneous~State}
In {\itshape Result 2}, we provide the correlation functional form of the constrained grand potential $\Omega_{\bm{v}}^{(m)}(\widetilde{G})$.
Section~\ref{functional integrals} will sketch how the perturbative field theory developed in Appendix \ref{sc} yields the correlation functional given in {\itshape Result 2}. 

Let us consider the inhomogeneous system characterized by the mean-field density $\rho_a^*(\bm{r})$ satisfying
\begin{flalign}
\label{results mean}
\rho_a^*(\bm{r})=e^{\beta\mu-\frac{c_{aa}({\bf 0})}{2}}\exp\left\{
\sum_{b=1}^m\int d\bm{r}'c_{ab}(\bm{r}-\bm{r}')\rho_b^*(\bm{r}')
\right\},
\end{flalign}
where $c_{ab}(\bm{r})$ denotes the two-point DCF (simply called DCF) between replica $a$ and replica $b$.
Here we suppose that a given function $\widetilde{G}(\bm{r},\bm{r}')$ imposed on the inter-replica correlation between replica $1$ and replica $a$ ($a\geq 2$) is expressed as
\begin{flalign}
\widetilde{G}(\bm{r},\bm{r}')=\rho_1^*(\bm{r})\rho_a^*(\bm{r}')\,\widetilde{g}(\bm{r}-\bm{r}')
=\rho_1^*(\bm{r})\rho_a^*(\bm{r}')\,\left\{1+\widetilde{h}(\bm{r}-\bm{r}')\right\},
\label{G g}
\end{flalign}
using a statistical realization of inter-replica radial distribution function $\widetilde{g}(\bm{r})$ or inter-replica total correlation function (TCF) $\widetilde{h}(\bm{r})\equiv \widetilde{g}(\bm{r})-1$ \cite{ls1,ls2}.
Namely, Equations~(\ref{identity3}) and (\ref{G g}) imply the constraint,
\begin{flalign}
\rho_1^*(\bm{r})\rho_a^*(\bm{r}')\left\{
1+\widetilde{h}(\bm{r}-\bm{r}')\right\}
=\rho_1(\bm{r})\rho_a(\bm{r}'),
\label{constraint2}
\end{flalign}
on $\rho_1(\bm{r})\rho_a(\bm{r}')$ which is a statistical realization of density-density correlation~\cite{ls1,ls2} as mentioned above.
Equation~(\ref{constraint2}) includes the trivial inter-replica constraints as follows: one constraint, $\overline{\rho}^2=\rho_1(\bm{r})\rho_a(\bm{r}')$ (i.e., $\widetilde{h}(\bm{r}-\bm{r}')=0$), forces the two-replica system to maintain uniformity without inter-replica correlations, whereas another constraint, $0=\rho_1(\bm{r})\rho_a(\bm{r}')$ (i.e., $\widetilde{h}(\bm{r}-\bm{r}')=-1$), imposes a region where two particles of different replicas exclude each other.
In Section~\ref{section result3}, we will see that the metastable TCF $\widetilde{h}_*(\bm{r}-\bm{r}')$ corresponds to the TCF obtained from averaging over statistical realizations of instantaneous density-density correlation $\rho_1(\bm{r})\rho_a(\bm{r}')$ consistently with Equation~(\ref{constraint2}) as well as the liquid-state theory~\cite{ls1,ls2}. 

Let $\bm{h}(\bm{r})$ and $\bm{c}(\bm{r})$ be the correlation matrices of TCFs and DCFs, respectively.
The intra-and inter-replica matrix elements vary, depending on whether replica 1 is included or not:
when setting $\bm{\chi}(\bm{r})=\bm{h}(\bm{r})$ or $\bm{c}(\bm{r})$ with the subscripts of their matrix elements denoting a pair of replicas, $\chi_{11}(\bm{r})=\chi_1(\bm{r})$ and $\chi_{aa}(\bm{r})=\chi(\bm{r})$ for $a\geq 2$, whereas $\chi_{1a}(\bm{r})=\chi_{a1}(\bm{r})=\widetilde{\chi}(\bm{r})$ for $a\geq 2$ and $\chi_{ab}(\bm{r})=\widetilde{\chi}\,'(\bm{r})$ for $a\neq b$ and $a,b\geq 2$.
As a consequence, we see from Equation~(\ref{results mean}) that
\begin{flalign}
\rho_1^*(\bm{r})=e^{\beta\mu-\frac{c_1({\bf 0})}{2}}\exp\left\{
\int d\bm{r}'c_1(\bm{r}-\bm{r}')\rho_1^*(\bm{r}')
+(m-1)\int d\bm{r}'\widetilde{c}(\bm{r}-\bm{r}')\rho^*(\bm{r}')
\right\},
\label{rho m}
\end{flalign}
where $\rho_a^*(\bm{r})=\rho^*(\bm{r})$ for $a\geq2$.
It is noted that the metastable density distribution $\rho_1^*(\bm{r})$ reduces to that from the Ramakrishnan-Yussouf density functional~\cite{dft1,dft2,dft3,ry}:
\begin{flalign}
\rho_1^*(\bm{r})&=e^{\beta\mu-\frac{c_1({\bf 0})}{2}}\exp\left\{
\int d\bm{r}'c_1(\bm{r}-\bm{r}')\rho_1^*(\bm{r}')
\right\},
\label{ry2}
\end{flalign}
in the limit of $m\rightarrow 1$.

The variational approach presented in Section~\ref{gb} justifies the following set of inhomogeneous Ornstein-Zernike equations~\cite{mor1,mor2,mor3,mor4,mor5,mor6,ls1,ls2}:
in general, we have
\begin{flalign}
h_{ac}(\bm{r}-\bm{r}')=c_{ac}(\bm{r}-\bm{r}')+\sum_{b=1}^m\int d\bm{r}'
\rho_{b}^*(\bm{r}")c_{ab}(\bm{r}-\bm{r}")h_{bc}(\bm{r}"-\bm{r}'),
\label{general oz}
\end{flalign}
which reads
\begin{flalign}
h_1(\bm{r}-\bm{r}')=c_1(\bm{r}-\bm{r}')+\int d\bm{r}"\rho_1^*(\bm{r}")
c_1(\bm{r}-\bm{r}")h_1(\bm{r}"-\bm{r}')\nonumber\\
\label{results oz intra}
+(m-1)\int d\bm{r}"\rho^*(\bm{r}")
\widetilde{c}(\bm{r}-\bm{r}")\widetilde{h}(\bm{r}"-\bm{r}'),
\end{flalign}
and

\vspace{-6pt}
\begingroup\makeatletter\def\f@size{9}\check@mathfonts
\def\maketag@@@#1{\hbox{\m@th\normalsize\normalfont#1}}
\begin{flalign}
\widetilde{h}(\bm{r}-\bm{r}')=\widetilde{c}(\bm{r}-\bm{r}')
&+\int d\bm{r}"\left\{
\rho^*(\bm{r}")\widetilde{c}(\bm{r}-\bm{r}")h(\bm{r}"-\bm{r}')
+\rho_1^*(\bm{r}")c_1(\bm{r}-\bm{r}")\widetilde{h}(\bm{r}"-\bm{r}')
\right\}\nonumber\\
\label{results oz inter}
&+(m-2)\int d\bm{r}"\rho^*(\bm{r}")
\widetilde{c}(\bm{r}-\bm{r}")\widetilde{h}(\bm{r}"-\bm{r}'),
\end{flalign}
\endgroup
in agreement with previous expressions~\cite{mor1,mor2,mor3,mor4,mor5,mor6}.

The second result ({\itshape Result 2}) can be obtained using the perturbative field theory at strong coupling (see Appendix \ref{sc}).
It will be shown in Section~\ref{functional integrals} that the constrained grand potential is of the following functional form:
\vspace{-6pt}
\begingroup\makeatletter\def\f@size{9}\check@mathfonts
\def\maketag@@@#1{\hbox{\m@th\normalsize\normalfont#1}}
\begin{flalign}
&\beta\Omega_{\bm{v}'}^{(m)}(\widetilde{G})
\nonumber\\
&\qquad
=\frac{1}{2}\int\!\!\!\int d\bm{r}_0d\bm{r}\,
\left\{
\rho_1^*(\bm{r}_0)\rho_1^*(\bm{r}_0-\bm{r})
+(m-1)\rho^*(\bm{r}_0)\rho^*(\bm{r}_0-\bm{r})
\right\}
g(\bm{r})v(\bm{r})
\nonumber\\
&\qquad+(m-1)\int\!\!\!\int d\bm{r}_0d\bm{r}\,
\rho_1^*(\bm{r}_0)\rho^*(\bm{r}_0-\bm{r})\widetilde{g}(\bm{r})\widetilde{v}(\bm{r})
\nonumber\\
&\qquad+\int d\bm{r}_0\,\left[\,\rho_1^*(\bm{r}_0)\left\{
\ln\rho_1^*(\bm{r}_0)-1-\beta\mu\right\}
+(m-1)\rho^*(\bm{r}_0)\left\{
\ln\rho^*(\bm{r}_0)-1-\beta\mu\right\}
\right]
\nonumber\\
&\qquad+\frac{1}{2}\int\!\!\!\int d\bm{r}_0d\bm{r}\,
\left\{\rho_1^*(\bm{r}_0)h_1(\bm{r})\delta(\bm{r})+(m-1)\rho^*(\bm{r}_0)h(\bm{r})\delta(\bm{r})
-\ln\left|\bm{S}\right|\right\}
\nonumber\\
\label{results omega}
&\qquad+(m-1)\int\!\!\!\int d\bm{r}_0d\bm{r}\,
\rho_1^*(\bm{r}_0)\rho^*(\bm{r}_0-\bm{r})\left\{
\widetilde{g}(\bm{r})\ln \widetilde{g}(\bm{r})-\widetilde{h}(\bm{r})-\widetilde{h}^2(\bm{r})+e^{\widetilde{h}(\bm{r})}-\widetilde{g}(\bm{r})
\right\},
\end{flalign}
\endgroup
where the matrix elements of $\bm{v}'$ has a non-zero potential $v_{1a}(\bm{r})=v_{a1}(\bm{r})=\widetilde{v}(\bm{r})$ between replica 1 and replica $a$ that enforces Equation~(\ref{constraint2}) without the constraint $\Delta_a(\widetilde{G},\boldsymbol{\widehat{\rho}})$, and~the matrix element of $\bm{S}$ is given by $S_{ab}(\bm{r})=\delta_{ab}\delta(\bm{r})+\rho_a^*(\bm{r}_0)h_{ab}(\bm{r})$.
It is noted that the last line of Equation~(\ref{results omega}) is reduced to the RHNC functional of $\widetilde{h}(\bm{r})$ in the approximation of $e^{\widetilde{h}(\bm{r})}-\widetilde{g}(\bm{r})\approx \widetilde{h}^2(\bm{r})/2$ \cite{mor1,mor2,mor3,mor4,mor5,mor6,ls1,ls2,morita,f2,f1}.

\subsection{{\itshape Result~3}: New Closure Obtained from the Generalized FP Potential}\label{section result3}
The stationary condition of $W(\widetilde{G})$ given by Equation~(\ref{results replica}) can be written as
\begin{flalign}
\label{results stationary}
\left.
\frac{\delta W(\widetilde{G})}{\delta\widetilde{h}}
\right|_{\widetilde{h}=\widetilde{h}_*}
=\frac{1}{N^*}\left.
\lim_{\widetilde{v}\rightarrow 0}
\frac{\delta}{\delta\widetilde{h}}
\left\{
\lim_{m\rightarrow 1}\frac{\partial\Omega_{\bm{v}'}^{(m)}(\widetilde{G})}{\partial m}
\right\}
\right|_{\widetilde{h}=\widetilde{h}_*}=0.
\end{flalign}
It is found from Equation~(\ref{results omega}) that
\vspace{-6pt}
\begingroup\makeatletter\def\f@size{9}\check@mathfonts
\def\maketag@@@#1{\hbox{\m@th\normalsize\normalfont#1}}
\begin{flalign}
\lim_{m\rightarrow 1}\frac{\partial\beta\Omega_{\bm{v}'}^{(m)}(\widetilde{G})}{\partial m}
&=\int\!\!\!\int d\bm{r}_0d\bm{r}\,
\left\{\frac{1}{2}\rho^*(\bm{r}_0)\rho^*(\bm{r}_0-\bm{r})g(\bm{r})v(\bm{r})
+\rho_1^*(\bm{r}_0)\rho^*(\bm{r}_0-\bm{r})\widetilde{g}(\bm{r})\widetilde{v}(\bm{r})\right\}
\nonumber\\
&+\int d\bm{r}_0\,\rho^*(\bm{r}_0)\left\{
\ln\rho^*(\bm{r}_0)-1-\beta\mu\right\}
+\frac{1}{2}\int\!\!\!\int d\bm{r}_0d\bm{r}\,
\rho^*(\bm{r}_0)h(\bm{r})\delta(\bm{r})
\nonumber\\
&+\frac{1}{2}\int\!\!\!\int
 d\bm{r}_0d\bm{r}\,\rho_1^*(\bm{r}_0)\rho^*(\bm{r}_0-\bm{r})
\widetilde{c}(\bm{r})\widetilde{h}(\bm{r})
\nonumber\\
\label{partial omega}
&+\int\!\!\!\int d\bm{r}_0d\bm{r}\,
\rho_1^*(\bm{r}_0)\rho^*(\bm{r}_0-\bm{r})\left\{
\widetilde{g}(\bm{r})\ln \widetilde{g}(\bm{r})-\widetilde{h}(\bm{r})-\widetilde{h}^2(\bm{r})+e^{\widetilde{h}(\bm{r})}-\widetilde{g}(\bm{r})
\right\},
\end{flalign}
\endgroup
where the third line of Equation~(\ref{partial omega}) is obtained from the derivative of the logarithmic term in the fifth line of Equation~(\ref{results omega}) with respect to $m$ using the Laplace expansion of $\left|\bm{S}\right|$ along the first row as follows:
\begin{flalign}
-\frac{1}{2}\frac{\partial}{\partial m}\ln\left|\bm{S}(\bm{r})\right|
&=-\frac{1}{2\left|\bm{S}(\bm{r})\right|}\left(
\frac{\partial\left|\bm{S}(\bm{r})\right|}{\partial m}
\right)
\nonumber\\
&=\frac{1}{2}\rho_1^*(\bm{r}_0)\rho^*(\bm{r}_0-\bm{r})
\widetilde{c}(\bm{r})\widetilde{h}(\bm{r}),
\label{partial s}
\end{flalign}
where use has been made of the cofactor expansion in calculating $\partial\left|\bm{S}(\bm{r})\right|/\partial m$.

It follows from Equation~(\ref{partial omega}) that the stationary condition (\ref{results stationary}) becomes
\vspace{-6pt}
\begingroup\makeatletter\def\f@size{9}\check@mathfonts
\def\maketag@@@#1{\hbox{\m@th\normalsize\normalfont#1}}
\begin{flalign}
\left.\frac{\delta W(\widetilde{G})}{\delta\widetilde{h}}
\right|_{\widetilde{h}=\widetilde{h}_*}
\approx\frac{1}{N^*}\int d\bm{r}_0\,
\rho^*(\bm{r}_0)\rho^*(\bm{r}_0-\bm{r})\left\{
\widetilde{c}_*(\bm{r})+\ln\widetilde{g}_*(\bm{r})-1-2\widetilde{h}_*(\bm{r})+e^{\widetilde{h}_*(\bm{r})}
\right\}=0,
\label{h derivative}
\end{flalign}
\endgroup
where the subscript 1 has been dropped because of the indistinguishability of all replicas in the limits of $m\rightarrow 1$ and $\widetilde{v}(\bm{r})\rightarrow 0$, $\delta\widetilde{h}/\delta\rho^*$ and its inverse are ignored, and~the first term on the rhs is an approximate form obtained from the third line of Equation~(\ref{partial omega}) (see \mbox{Appendix \ref{appendix ln}} for the detailed derivation).
We can easily verify the equivalence between Equation~(\ref{h derivative}) and the following closure:
\begin{flalign}
\label{results closure}
\widetilde{g}_*(\bm{r})&=e^{\widetilde{h}_*(\bm{r})-\widetilde{c}_*(\bm{r})+B(\bm{r})},
\\
\label{results bridge}
B(\bm{r})&=\widetilde{g}_*(\bm{r})-e^{\widetilde{h}_*(\bm{r})},
\end{flalign}
which corresponds to the third result ({\itshape Result 3}), a~new closure in the context of the liquid-state theory~\cite{ls1,ls2}.

Two remarks on Equations~(\ref{results stationary}), (\ref{results closure}) and (\ref{results bridge}) are in~order:
\begin{itemize}
\item Equation~(\ref{results stationary}) is valid when a metastable state at $\widetilde{h}_*(\bm{r})=\widetilde{h}(\bm{r})$ is stable in the vanishing limit of the inter-replica interaction potential (i.e., $\widetilde{v}(\bm{r})\rightarrow 0$); otherwise, transitions between basins occur in the FEL and the inter-replica correlations disappear, thereby amounting to $\widetilde{g}_*(\bm{r})=1+\widetilde{h}_*(\bm{r})=1$, the~trivial solution to Equation~(\ref{results stationary}). In~other words, the~new closure (\ref{results closure}) applies to the metastable state defined by Equation~(\ref{results stationary}).
\item The bridge function $B(\bm{r})$ given by Equation~(\ref{results bridge}) is approximated by $B(\bm{r})=-\widetilde{h}_*^2(\bm{r})/2$, which coincides with the main term of either the soft mean spherical approximation (MSA) or various approximations used for hard-sphere systems~\cite{ls2,f2}.
\end{itemize}

\subsection{{\itshape Result~4}: Self-Consistent Equation for the Non-Ergodicity Parameter}
In the fourth result ({\itshape Result 4}), we restrict ourselves to uniform systems in Fourier space.
We introduce the non-ergodicity parameter $f(\bm{k})$ by relating the inter-replica TCF $\widetilde{h}_*(\bm{k})$ to the intra-replica structure factor $S(\bm{k})=1+\overline{\rho}h_*(\bm{k})$ \cite{mct1,mct2,mor1,mor2,mor3,mor4,mor5,mor6}:
\begin{flalign}
\label{results def non-ergodicity}
f(\bm{k})=\frac{\overline{\rho}\widetilde{h}_*(\bm{k})}{S(\bm{k})}.
\end{flalign}
We need to find an approximation of the closure (\ref{results closure}) that is available to obtain the self-consistent equation including terms up to quadratic order in the non-ergodicity parameter $f(\bm{k})$.
It is appropriate for this purpose to expand the rhs of the closure (\ref{results closure}), providing
\begin{flalign}
\widetilde{g}_*(\bm{r})\approx
\widetilde{g}_*(\bm{r})-\widetilde{c}_*(\bm{r})+B(\bm{r})+\frac{1}{2}
\left\{\widetilde{h}_*(\bm{r})-\widetilde{c}_*(\bm{r})\right\}^2.
\label{approx closure}
\end{flalign}
Equation~(\ref{approx closure}) reads in Fourier space 
\begin{flalign}
\widetilde{c}_*(\bm{k})=\frac{1}{2}\int d\bm{q}\,
\left\{
\widetilde{c}_*(\bm{q})\,\widetilde{c}_*(\bm{k}-\bm{q})-\widetilde{c}_*(\bm{q})\,\widetilde{h}_*(\bm{k}-\bm{q})
-\widetilde{h}_*(\bm{q})\,\widetilde{c}_*(\bm{k}-\bm{q})
\right\},
\label{approx closure2}
\end{flalign}
when making the approximation of $B(\bm{r})\approx -h_*^2(\bm{r})/2$ as remarked after Equation~(\ref{results bridge}).
Meanwhile, the~neglect of inhomogeneity (i.e., $\rho^*(\bm{r})=\overline{\rho}$) allows us to express the Fourier transform of the Ornstein-Zernike Equation~(\ref{results oz inter}) at $m=1$ as
\begin{flalign}
\widetilde{c}_*(\bm{k})=\frac{1}{\overline{\rho}S(\bm{k})}\left\{\frac{f(\bm{k})}{1-f(\bm{k})}\right\},
\label{fourier oz}
\end{flalign}
using the non-ergodicity parameter $f(\bm{k})$ defined by Equation~(\ref{results def non-ergodicity}).

Combining Equations~(\ref{results def non-ergodicity}), (\ref{approx closure2}) and (\ref{fourier oz}), we obtain the self-consistent equation for $f(\bm{k})$ ({\itshape Result 4}):
\begin{flalign}
\frac{f(\bm{k})}{1-f(\bm{k})}
=\frac{S(\bm{k})}{2\overline{\rho}}\int d\bm{q}\,
\mathcal{M}(\bm{k},\bm{q})\,S(\bm{q})\,S(\bm{k}-\bm{q})
\,f(\bm{q})\,f(\bm{k}-\bm{q})+\mathcal{O}[f^3],
\label{scf}
\end{flalign}
where the inverse of the intra-replica structure factor $S(\bm{q})$ is related to the intra-replica DCF $c_*(\bm{q})$ as $1/S(\bm{q})=1-\overline{\rho}c_*(\bm{q})$ and the kernel $\mathcal{M}(\bm{k},\bm{q})$ is given by
\begin{flalign}
\mathcal{M}(\bm{k},\bm{q})
&=\frac{1}{S^2(\bm{q})\,S^2(\bm{k}-\bm{q})}
-\frac{1}{S^2(\bm{q})}
-\frac{1}{S^2(\bm{k}-\bm{q})}
\nonumber\\
&=\left\{\overline{\rho}^2c_*(\bm{q})\,c_*(\bm{k}-\bm{q})\right\}^2
+2\overline{\rho}^2c_*(\bm{q})\,c_*(\bm{k}-\bm{q})\left\{
\frac{1}{S(\bm{q})}
+\frac{1}{S(\bm{k}-\bm{q})}
\right\}-1;
\label{kernel s}
\end{flalign}
see Appendix \ref{appendix kernel} for details.
We can relate the product $c_*(\bm{q})\,c_*(\bm{k}-\bm{q})$ in Equation~(\ref{kernel s}) to the triplet DCF $c_*^{(3)}(\bm{q},\bm{k}-\bm{q})$ by adopting the approximate form as follows:
\begin{flalign}
c_*^{(3)}(\bm{q},\bm{k}-\bm{q})=\frac{c_*^{(3)}(\bm{0},\bm{0})}{\left\{c_*(\bm{0})\right\}^2}\,c_*(\bm{q})\,c_*(\bm{k}-\bm{q}),
\label{c3}
\end{flalign}
which is validated by the weighted density approximation or the closure-based density functional theory~\cite{dcf1,dcf2,dcf3}.
The expression (\ref{c3}) and the introduction of the negative factor, $\alpha=\left\{c_*(\bm{0})\right\}^2/c_*^{(3)}(\bm{0},\bm{0})<0$, transform Equation~(\ref{kernel s}) into the following kernel ({\itshape Result 4}):
\begin{flalign}
\label{results kernel}
\mathcal{M}(\bm{k},\bm{q})
=\left\{\overline{\rho}^2\alpha c_*^{(3)}(\bm{q},\bm{k}-\bm{q})\right\}^2
+2\overline{\rho}^2\alpha c_*^{(3)}(\bm{q},\bm{k}-\bm{q})\left\{
\frac{1}{S(\bm{q})}
+\frac{1}{S(\bm{k}-\bm{q})}
\right\}-1,
\end{flalign}
where $\alpha=\left\{c_*(\bm{0})\right\}^2/c_*^{(3)}(\bm{0},\bm{0})$ and $c_*^{(3)}(\bm{q},\bm{k}-\bm{q})$ denotes the triplet DCF~\cite{dcf1,dcf2,dcf3}.
It is noted that Equation~(\ref{results kernel}) can be compared with the previous result from other static theories~\cite{mor5,dfr4,dfr5}:
the systematic expansion methods lead to the appearance of the triplet DCF in the kernel~\cite{mor5,dfr4,dfr5}, similar to Equation~(\ref{results kernel}).

\section{Derivation process of {\itshape Result~2}}\label{functional integrals}
This section presents a scheme to obtain {\itshape Result 2} based on the strong-coupling perturbation theory (see Appendix \ref{sc}).
To this end, we focus on how to perform the functional integrals over one-body and two-body potential fields appearing in Equations~(\ref{delta omega approximation2}), (\ref{average nu}), (\ref{def mayer f}) and (\ref{f0 integral})--(\ref{phi average}).
\subsection{One-Body Potential Field (1): Evaluating Equation~(\ref{f0 integral}) in the Saddle-Point Approximation}
We see from Equation~(\ref{def a}) that the saddle-point equation $\left.\delta\mathcal{H}_{\mathrm{mf}}(\boldsymbol{\phi})/\delta\phi\right|_{\phi_a=i\psi_a^*}=0$ in Equation~(\ref{f0 integral}) gives
\begin{flalign}
\label{saddle eq}
\left.
\frac{\delta\beta\mathcal{H}_0(\bm{c},\boldsymbol{\phi})}{\delta\phi_a}
\right|_{\phi_a=i\psi_a^*}
=\left(\frac{\overline{\rho}}{\gamma}\right)
\left.
\frac{\delta\mathcal{U}_1(\boldsymbol{\phi})}{\delta\phi_a}
\right|_{\phi_a=i\psi_a^*}.
\end{flalign}
Substituting Equations~(\ref{def h0}) and (\ref{def u1}) into Equation~(\ref{saddle eq}), we have
\begin{flalign}
\psi_a^*(\bm{r})&=\frac{c_{aa}({\bf 0})}{2}-e^{\beta\mu}
\sum_{b=1}^m\int d\bm{r}'c_{ab}(\bm{r}-\bm{r}')
e^{-\psi_b^*(\bm{r}')}.
\label{saddle psi}
\end{flalign}
We can verify that Equation~(\ref{saddle psi}) transforms to Equation~(\ref{results mean}) by setting $\rho_a^*(\bm{r})=e^{\beta\mu-\psi_a^*(\bm{r})}$.

Let $F_{\mathrm{mf}}(-k_BT\bm{c},\boldsymbol{\rho}^*)$ be the mean-field free energy defined by
\begin{flalign}
F_{\mathrm{mf}}(-k_BT\bm{c},\boldsymbol{\rho}^*)
=U(-k_BT\bm{c},\boldsymbol{\rho}^*)-T\mathcal{S}^{\mathrm{id}}(\boldsymbol{\rho}^*),
\label{def mf}
\end{flalign}
where $U(\bm{v},\boldsymbol{\widehat{\rho}})$ has been defined in Equation~(\ref{results def u}) and $\mathcal{S}^{\mathrm{id}}(\boldsymbol{\rho}^*)$ denotes the sum of ideal gas entropy $\mathcal{S}_a^{\mathrm{id}}(\rho_a^*)$ given by Equation~(\ref{def entropy a}):
\vspace{-6pt}
\begingroup\makeatletter\def\f@size{9}\check@mathfonts
\def\maketag@@@#1{\hbox{\m@th\normalsize\normalfont#1}}
\begin{flalign}
-T\mathcal{S}^{\mathrm{id}}(\boldsymbol{\rho}^*)
&=-T\sum_{a=1}^m \mathcal{S}_a^{\mathrm{id}}(\rho_a^*)
\nonumber\\
&=k_BT\int d\bm{r}_0\,\left[\,\rho_1^*(\bm{r}_0)\left\{
\ln\rho_1^*(\bm{r}_0)-1-\beta\mu\right\}
+(m-1)\rho^*(\bm{r}_0)\left\{
\ln\rho^*(\bm{r}_0)-1-\beta\mu\right\}
\right].
\label{def entropy}
\end{flalign}
\endgroup
Plugging Equation~(\ref{results mean}) into Equations~(\ref{def mf}) and (\ref{def entropy}), we find
\begin{flalign}
\beta F_{\mathrm{mf}}(-k_BT\bm{c},\boldsymbol{\rho}^*)
&=\frac{1}{2}
\int\!\!\!\int d\bm{r}d\bm{r}'\boldsymbol{\rho}^*(\bm{r})^{\rm{T}}\bm{c}(\bm{r}-\bm{r}')
\boldsymbol{\rho}^*(\bm{r}')
-\sum_{a=1}^m\int d\bm{r}\,\rho_a^*(\bm{r})\nonumber\\
&=\beta\mathcal{H}_{\mathrm{mf}}(i\boldsymbol{\psi}^*)
\label{saddle total}
\end{flalign}
(see also Appendix \ref{appendix mf} for details of the last equality).

The quadratic terms due to fluctuations around the saddle-point path $i\boldsymbol{\psi}^*$ are written as
\begin{flalign}
&\beta\mathcal{H}_{\mathrm{mf}}(\boldsymbol{\varphi}+i\boldsymbol{\psi}^*)-\beta\mathcal{H}_{\mathrm{mf}}(i\boldsymbol{\psi}^*)
\nonumber\\
&\qquad\approx
-\frac{1}{2}\int\!\!\!\int d\bm{r}d\bm{r}'
\boldsymbol{\varphi}^{\rm{T}}(\bm{r})\bm{c}^{-1}(\bm{r}-\bm{r}')\boldsymbol{\varphi}(\bm{r}')
+\sum_{a=1}^m\frac{1}{2}\int d\bm{r}\rho_a^*(\bm{r})\varphi_a^2(\bm{r})
\nonumber\\
&\qquad=-\frac{1}{2}
\int\!\!\!\int d\bm{r}d\bm{r}'\boldsymbol{\varphi}(\bm{r})^{\rm{T}}\bm{h}^{-1}(\bm{r}-\bm{r}')
\boldsymbol{\varphi}(\bm{r}').
\label{quad exp}
\end{flalign}
In the last equality of Equation~(\ref{quad exp}), use has been made of the following relation:
\begin{flalign}
\label{h-1c-1}
h_{ab}^{-1}(\bm{r}-\bm{r}')
=c_{ab}^{-1}(\bm{r}-\bm{r}')-\rho^*_a(\bm{r})\delta_{ab}\delta(\bm{r}-\bm{r}'),
\end{flalign}
which is equivalent to the inhomogeneous Ornstein-Zernike Equations~(\ref{general oz}) as confirmed in Appendix \ref{appendix inverse oz}.
It is found from Equations~(\ref{saddle total}) and (\ref{quad exp}) that the saddle-point approximation of Equation~(\ref{f0 integral}) yields
\begin{flalign}
e^{-\beta\mathcal{F}(\boldsymbol{\nu}=\bm{0})}
&=\frac{1}{\mathcal{N}_c}\,e^{-\beta F_{\mathrm{mf}}(-k_BT\bm{c},\boldsymbol{\rho}^*)}
\int D\boldsymbol{\varphi}\,
e^{\frac{1}{2}
\int\!\!\!\int d\bm{r}d\bm{r}'\boldsymbol{\varphi}(\bm{r})^{\rm{T}}\bm{h}^{-1}(\bm{r}-\bm{r}')
\boldsymbol{\varphi}(\bm{r}')}.
\label{sp0}
\end{flalign}
Equations~(\ref{def norm c})--(\ref{zero}) further imply that Equation~(\ref{sp0}) is transformed into
\begin{flalign}
e^{-\beta\mathcal{F}(\boldsymbol{\nu}=\bm{0})}
&=\frac{\mathcal{N}_h}{\mathcal{N}_c}\,e^{-\beta F_{\mathrm{mf}}(-k_BT\bm{c},\boldsymbol{\rho}^*)}
\nonumber\\
&=\frac{1}{\mathcal{N}_c}\,e^{-\beta F_{\mathrm{mf}}(-k_BT\bm{c},\boldsymbol{\rho}^*)}
\int D\boldsymbol{\varphi}\,
e^{-\beta\mathcal{H}_0(\bm{h},\boldsymbol{\varphi})}.
\label{sp1}
\end{flalign}
We will use the last line on the rhs of Equation~(\ref{sp1}) as a reference form in evaluating $\beta\mathcal{F}(\boldsymbol{\nu})-\beta\mathcal{F}(\boldsymbol{\nu}=\bm{0})$ given by Equation~(\ref{sc main}).

It follows from Equations~(\ref{f0 integral}) and (\ref{sp1}) that
\begin{flalign}
\beta\Phi_{-k_BT\bm{c}}^{(m)}
=\beta F_{\mathrm{mf}}(-k_BT\bm{c},\boldsymbol{\rho}^*)
-\ln\frac{\mathcal{N}_h}{\mathcal{N}_c},
\label{phi c}
\end{flalign}
where $\mathcal{N}_h/\mathcal{N}_c$ is related to the determinant of the matrix, $\bm{S}=\bm{c}^{-1}\bm{h}$, as~\begin{flalign}
\frac{\mathcal{N}_h}{\mathcal{N}_c}=\left\{\prod_{\bm{r},\bm{r}'}\left|\bm{S}(\bm{r}-\bm{r}')\right|\right\}^{1/2},
\label{nc nh}
\end{flalign}
and the matrix element of $\bm{S}$ is given by
\vspace{-6pt}
\begingroup\makeatletter\def\f@size{9}\check@mathfonts
\def\maketag@@@#1{\hbox{\m@th\normalsize\normalfont#1}}
\begin{flalign}
&S_{ac}(\bm{r}-\bm{r}')\equiv\sum_{b=1}^m\int d\bm{r}"\,c_{ab}^{-1}(\bm{r}-\bm{r}")h_{bc}(\bm{r}"-\bm{r}')
\nonumber\\
&=\sum_{b=1}^m\int d\bm{r}"\left\{
c_{ab}^{-1}(\bm{r}-\bm{r}")c_{bc}(\bm{r}"-\bm{r}')
+\sum_{d=1}^m\int d\bm{u}\,
\rho_d^*(\bm{u})c_{ab}^{-1}(\bm{r}-\bm{r}")c_{bd}(\bm{r}"-\bm{u})h_{dc}(\bm{u}-\bm{r}')
\right\}
\nonumber\\
&=\delta_{ac}\delta(\bm{r}-\bm{r}')
+\sum_{d=1}^m\int d\bm{u}\,\rho_d^*(\bm{u})\delta_{ad}\delta(\bm{r}-\bm{u})h_{dc}(\bm{u}-\bm{r}')
\nonumber\\
&=\delta_{ac}\delta(\bm{r}-\bm{r}')
+\rho_a^*(\bm{r})h_{ac}(\bm{r}-\bm{r}')
\nonumber\\
&=\frac{\left<\rho_a(\bm{r})\rho_c(\bm{r}')\right>}{\rho_c^*(\bm{r}')}\geq 0,
\label{c-1 h}
\end{flalign}
\endgroup
ensuring that $|\bm{S}|=|\bm{c}^{-1}\bm{h}|\geq 0$.
Replacing $\bm{r}$ and $\bm{r}'$ by $\bm{r}_0$ and $\bm{r}_0-\bm{r}$, respectively, in \mbox{Equation~(\ref{c-1 h})}, we have
\begin{flalign}
-\ln\frac{\mathcal{N}_h}{\mathcal{N}_c}
=-\frac{1}{2}\int\!\!\!\int d\bm{r}_0d\bm{r}\,
\ln\left|\bm{S}\right|,
\label{ln}
\end{flalign}
in agreement with the logarithmic term in Equation~(\ref{results omega}).

\subsection{One-Body Potential Field (2): Perturbative Calculation of Equation~(\ref{delta omega approximation2})}\label{sc calculation}
Remembering that $\rho_a^*(\bm{r})=e^{\beta\mu-\psi_a^*(\bm{r})}$, the~average term in Equation~(\ref{sc main}) becomes
\begin{flalign}
&\left(\frac{\overline{\rho}}{\gamma}\right)^2
\,\left<
e^{\int d\bm{r}\,
\left\{i \phi_1(\bm{r})\widehat{\rho}_1^{(1)}(\bm{r})
+i \phi_a(\bm{r})\widehat{\rho}_a^{(1)}(\bm{r})\right\}}
\right>_{\phi}\nonumber\\
&\qquad=\rho_1^*(\bm{r}_{1,1})\rho^*(\bm{r}_{a,1})
\,\left<
e^{\int d\bm{r}\,
\left\{i \varphi_1(\bm{r})\widehat{\rho}_1^{(1)}(\bm{r})
+i \varphi_a(\bm{r})\widehat{\rho}_a^{(1)}(\bm{r})\right\}}
\right>_{\varphi}
\nonumber\\
&\qquad=\rho_1^*(\bm{r}_{1,1})\rho^*(\bm{r}_{a,1})\,e^{\widetilde{h}(\bm{r}_{1,1}-\bm{r}_{a,1})},
\label{sc sp}
\end{flalign}
where the subscript $\varphi$ denotes the following average:
\begin{flalign}
\left<\mathcal{O}\right>_{\varphi}
=\frac{\int D\boldsymbol{\varphi}\,\mathcal{O}\,
e^{-\beta\mathcal{H}_0(\bm{h},\boldsymbol{\varphi})}}
{\int D\boldsymbol{\varphi}\,
e^{-\beta\mathcal{H}_0(\bm{h},\varphi)}},
\label{varphi average}
\end{flalign}
according to Equation~(\ref{sp1}) (see Appendix \ref{appendix sc} for the detailed derivation of Equation~(\ref{sc sp})).

It is noted that the one-particle densities, $\widehat{\rho}_1^{(1)}(\bm{r})$ and $\widehat{\rho}_a^{(1)}(\bm{r})$, of~replicas 1 and $a$ in \mbox{Equation~(\ref{sc sp})} represent the two-particle system as a mixture of two replicas.
Accordingly, the~last line on the rhs of Equation~(\ref{sc sp}) reduces to $\rho^*(\bm{r}_{a,1})\rho_1^*(\bm{r}_{1,1})$ in the absence of inter-replica correlation between two particles of different replicas (i.e., $\widetilde{h}(\bm{r}_{a,1}-\bm{r}_{1,1})=0$) consistently with the following result for the sum of one-particle systems:
\begin{flalign}
\left<\frac{\overline{\rho}}{\gamma}\mathcal{U}_1(\boldsymbol{\phi})\right>_{\phi}
=\sum_{a=1}^m\int d\bm{r}_{a,1}\,\rho_a^*(\bm{r}_{a,1})
\,\left<
e^{\int d\bm{r}\,i \varphi_a(\bm{r})\widehat{\rho}_a^{(1)}(\bm{r})}
\right>_{\varphi}
=\sum_{a=1}^m\int d\bm{r}_{a,1}\,\rho_a^*(\bm{r}_{a,1}),
\label{u1 result}
\end{flalign}
where the above $\varphi$-averaging is applied to the one-particle term $\mathcal{U}_1(\boldsymbol{\phi})$ given by Equation~(\ref{def u1}), or~setting $\mathcal{O}=(\overline{\rho}/\gamma)\,\mathcal{U}_1(\boldsymbol{\varphi}+i\boldsymbol{\psi}^*)$ in Equation~(\ref{varphi average}) because of $\boldsymbol{\phi}=\boldsymbol{\varphi}+i\boldsymbol{\psi}^*$.

Combining Equations~(\ref{delta omega approximation}), (\ref{functional delta}), (\ref{average nu}), (\ref{sc main}) and (\ref{sc sp}), we obtain the additional contribution to $\beta\Phi_{\bm{v}'}^{(m)}$ given by Equations~(\ref{maximum}), (\ref{def mf}), (\ref{phi c}) and (\ref{ln}):
\begin{flalign}
e^{-\beta\Omega_{\bm{v}'}^{(m)}(\widetilde{G})+\beta\Phi_{\bm{v}'}^{(m)}}
&=\left<\prod_{a=2}^m
\Delta_a(\widetilde{G},\rho)
\right>_{\bm{c}}
\nonumber\\
&=\int D'\boldsymbol{\nu}\,
e^{-\sum_{a=2}^m\Gamma_a(\boldsymbol{\nu})},
\label{diff phi int}
\end{flalign}
and
\begin{flalign}
\Gamma_a(\boldsymbol{\nu})=-\int\!\!\!\int d\bm{r}d\bm{r}'\,
\rho_1^*(\bm{r})\rho^*(\bm{r}')\left\{
i\,\widetilde{g}(\bm{r}-\bm{r}')\,\nu_a(\bm{r}-\bm{r}')
+e^{\widetilde{h}(\bm{r}-\bm{r}')}\,
f(i\,\nu_a)
\right\};
\label{gamma a}
\end{flalign}
see Equation~(\ref{delta omega approximation}) for the definition of $\left<\mathcal{O}\right>_{\bm{c}}$.
The results from the strong-coupling perturbation method developed in Appendix \ref{sc} are summed up in Equations~(\ref{diff phi int}) and (\ref{gamma a}).

\subsection{Two-Body Potential Field: Derivation of {\itshape Result~2}: Rearrangements in the Mean-Field Approximation of Equation~(\ref{diff phi int})}\label{rearrange}
There are two remaining steps toward obtaining Equation~(\ref{results omega}):
the first step is to evaluate the $\boldsymbol{\nu}$-functional integral given by Equation~(\ref{diff phi int}), and~the second step is to rearrange the interaction energy when adding the last two terms on the rhs of Equation~(\ref{maximum}) to $U(-k_BT\bm{c},\boldsymbol{\rho}^*)$.

First, let us evaluate the $\boldsymbol{\nu}$--field integral given by Equation~(\ref{diff phi int}) in the mean-field approximation.
Equations~(\ref{diff phi int}) and (\ref{gamma a}) provide the saddle-point equation as follows:
\begin{flalign}
\left.
\frac{\delta\Gamma_a(\boldsymbol{\nu})}{\delta\nu_a}
\right|_{\nu_a^*=iu}=0,
\label{sp nu}
\end{flalign}
giving
\begin{flalign}
\widetilde{g}(\bm{r})=e^{\widetilde{h}(\bm{r})+u(\bm{r})},
\label{sp nu2}
\end{flalign}
similar to a closure in the liquid-state theory~\cite{ls1,ls2} though given correlation functions do not necessarily satisfy any closure, other than the Ornstein-Zernike equation.
Substituting Equation~(\ref{sp nu2}) into Equation~(\ref{gamma a}), we obtain
\begin{flalign}
\Gamma_a(\boldsymbol{\nu}^*)
=\int\!\!\!\int d\bm{r}_0d\bm{r}\,
\rho_1^*(\bm{r}_0)\rho^*(\bm{r}_0-\bm{r})\left[\,
\widetilde{g}(\bm{r})
\left\{
\ln\widetilde{g}(\bm{r})-\widetilde{h}(\bm{r})
\right\}
+e^{\widetilde{h}(\bm{r})}
-\widetilde{g}(\bm{r})
\right],
\label{mf gamma}
\end{flalign}
or
\begin{flalign}
&\Omega_{\bm{v}'}^{(m)}(\widetilde{G})-\Phi_{\bm{v}'}^{(m)}
\nonumber\\
&\quad=(m-1)\int\!\!\!\int d\bm{r}_0d\bm{r}\,
\rho_1^*(\bm{r}_0)\rho^*(\bm{r}_0-\bm{r})\left[\,
\widetilde{g}(\bm{r})
\left\{
\ln\widetilde{g}(\bm{r})-\widetilde{h}(\bm{r})
\right\}
+e^{\widetilde{h}(\bm{r})}
-\widetilde{g}(\bm{r})
\right],
\label{diff functional}
\end{flalign}
due to Equations~(\ref{diff phi int}) and (\ref{gamma a}).

Next, we rewrite the interaction energy.
Considering the expression (\ref{results def u}) and the Ornstein-Zernike equation,
\begin{flalign}
h_{aa}(\bm{0})=c_{aa}(\bm{0})+\sum_{b=1}^m\int d\bm{r}'\rho_b^*(\bm{r}')h_{ab}(\bm{r}-\bm{r}')c_{ab}(\bm{r}-\bm{r}'),
\label{oz zero}
\end{flalign}
we have
\begin{flalign}
U(-k_BT\bm{c},\boldsymbol{\rho}^*)
&+\sum_{a=1}^m\sum_{b=1}^m\frac{1}{2}\int\!\!\!\int d\bm{r}d\bm{r}'
\rho_a^*(\bm{r})\rho_b^*(\bm{r}')g_{ab}(\bm{r}-\bm{r}')c_{ab}(\bm{r}-\bm{r}')
\nonumber\\
&=\frac{1}{2}\sum_{a=1}^m\int\!\!\!\int d\bm{r}_0d\bm{r}\,
\rho_a^*(\bm{r}_0)h_{aa}(\bm{r})\delta(\bm{r})
\nonumber\\
&=\frac{1}{2}\int\!\!\!\int d\bm{r}_0d\bm{r}\,
\left\{\rho_1^*(\bm{r}_0)h_1(\bm{r})\delta(\bm{r})+(m-1)\rho^*(\bm{r}_0)h(\bm{r})\delta(\bm{r})\right\}.
\label{u reduction1}
\end{flalign}
To clarify the difference between the bare interaction potentials of $\bm{v}$ and $\bm{v}'$, we also separate the intra-replica interaction term from the inter-replica one created by $\widetilde{v}(\bm{r})=v_{a1}(\bm{r})=v_{1a}(\bm{r})$: 
\begin{flalign}
\sum_{a=1}^m\sum_{b=1}^m\frac{1}{2}\int\!\!\!\int d\bm{r}_0d\bm{r}
&\rho_a^*(\bm{r})\rho_b^*(\bm{r}')g_{ab}(\bm{r}-\bm{r}')v_{ab}(\bm{r}-\bm{r}')
\nonumber\\
=&\frac{1}{2}\int\!\!\!\int d\bm{r}_0d\bm{r}\,
\left\{
\rho_1^*(\bm{r}_0)\rho_1^*(\bm{r}_0-\bm{r})
+(m-1)\rho^*(\bm{r}_0)\rho^*(\bm{r}_0-\bm{r})
\right\}
g(\bm{r})v(\bm{r})
\nonumber\\
&+(m-1)\int\!\!\!\int d\bm{r}_0d\bm{r}\,
\rho_1^*(\bm{r}_0)\rho^*(\bm{r}_0-\bm{r})
\widetilde{g}(\bm{r})\widetilde{v}(\bm{r}).
\label{u reduction2}
\end{flalign}
Combining Equations~(\ref{def mf}), (\ref{def entropy}), (\ref{phi c}), (\ref{ln}), (\ref{diff functional}), (\ref{u reduction1}) and (\ref{u reduction2}), we obtain $\beta\Omega_{\bm{v}'}^{(m)}(\widetilde{G})$ expressed by Equation~(\ref{results omega}), namely {\itshape Result 2}.

\section{Concluding~Remarks}\label{section con}
The generalized FP potential $W(\widetilde{G})$ as a functional of given TCF $\widetilde{h}(\bm{r})$ is similar to the original FP potential~\cite{fp1,fp2,fp3,fp4,fp5,fp6,fp7,fp8,fp9,fp10,fp11} in that both have constraints on inter-replica correlations.
The difference is that the generalized FP potential adopts a local order parameter instead of a global order one, the~overlap $Q$ (see Equation~(\ref{def q})), used in the original FP potential $V(Q)$.
Upon reviewing the formulation of $W(\widetilde{G})$ presented so far, we find two essentials for the field-theoretical achievements.
The former lies in the variational method described in Appendix \ref{gb}, whereas Equation~(\ref{delta omega approximation}) represents the latter.
The details~follow:
\begin{itemize}
\item {\itshape Unconstrained grand potential mimicking inter-replica correlations}
: At first, we consider a coupled $m$-replica system that reproduces a given distribution of the inter-replica TCF $\widetilde{h}(\bm{r})$ without constraints.
We tune the inter-replica interaction potential $\widetilde{v}(\bm{r})$ to mimic the inter-replica correlations.
From evaluating the free-energy functional without constraints in the Gaussian approximation, we obtain the same functional form as the random phase approximation (RPA) in terms of the liquid-state theory~\cite{ls1,ls2}; however, the~density distribution is different.
The variational method presented in Appendix \ref{gb} justifies the input of the density distribution given by Equation~(\ref{rho m}), which converges to that of the Ramakrishnan--Yussouf density functional theory~\cite{ry} in the limit of $m\rightarrow 1$ as demonstrated in Equation~(\ref{ry2}).
\item {\itshape Evaluating the difference between the constrained and unconstrained grand potentials}: Next, we take the free-energy functional of the unconstrained system as a reference energy.
Equation~(\ref{delta omega approximation}) indicates that the field-theoretical formulation focuses on the free energy difference between the constrained and unconstrained free-energy functionals.
The strong-coupling expansion method developed in Appendix \ref{sc section} allows us to evaluate this difference in Sections~\ref{sc calculation} and \ref{rearrange}.
Thus, we obtain Equation~(\ref{diff functional}), the~constraint-associated free energy difference as a functional of inter-replica TCF $\widetilde{h}(\bm{r})$ and density distribution $\rho^*(\bm{r})$ determined by the Ramakrishnan-Yussouf theory~\cite{ry}.
\end{itemize}

Equation~(\ref{diff functional}) reduces to the functional difference between the $\widetilde{h}(\bm{r})$-dependent parts in the HNC and RPA approximations when substituting $e^{\widetilde{h}(\bm{r})}-\widetilde{g}(\bm{r})\approx \widetilde{h}^2(\bm{r})/2$ into Equation~(\ref{diff functional}).
This agreement indicates consistency between the field-theoretical formalism in this paper and the Legendre-transform-based theory using the Morita-Hiroike functional~\cite{mor1,mor2,mor3,mor4,mor5,mor6,morita}.

Combination of Equations~(\ref{results replica}) and (\ref{partial omega}) gives the difference between the generalized FP potentials at zero and a finite value of the inter-replica TCFs as follows:
\begin{flalign}
&W\left(\widetilde{G}=\overline{\rho}^2\left\{1+\widetilde{h}\right\}\right)-W(\overline{\rho}^2)\nonumber\\
&\qquad=\frac{1}{2N^*}\int\!\!\!\int
 d\bm{r}_0d\bm{r}\,\rho^*(\bm{r}_0)\rho^*(\bm{r}_0-\bm{r})
\widetilde{c}(\bm{r})\widetilde{h}(\bm{r})
\nonumber\\
\label{potential diff}
&\qquad+\frac{1}{N^*}\int\!\!\!\int d\bm{r}_0d\bm{r}\,
\rho^*(\bm{r}_0)\rho^*(\bm{r}_0-\bm{r})\left\{
\widetilde{g}(\bm{r})\ln \widetilde{g}(\bm{r})-\widetilde{h}(\bm{r})-\widetilde{h}^2(\bm{r})+e^{\widetilde{h}(\bm{r})}-\widetilde{g}(\bm{r})
\right\}.
\end{flalign}
The potential difference in Equation~(\ref{potential diff}) arises from the entropic cost of localizing the system in an arbitrary state.
It is noted, however, that the closure given by Equations~(\ref{results closure}) and (\ref{results bridge}) applies only to Equation~(\ref{potential diff}) in a metastable state characterized by $\widetilde{h}_*(\bm{r})$, which is in contrast to the Morita-Hiroike functional covering only the inter-replica TCF that necessarily satisfies the conventional closure~\cite{ls1,ls2} of the liquid-state theory due to the Legendre-transform-based formalism.
That is, the~generalized FP potential expressed as Equation~(\ref{potential diff}) has a characteristic inherited from the original FP theory, a~Landau-type theory relevant to investigate the FEL.
Furthermore, Equation~(\ref{potential diff}) represents that our study provides the basis of Ginzburg--Landau-type theory~\cite{text} as an extension of Landau-type one:
the generalized FP potential $W(\widetilde{G})$ as a functional of local order parameter $\widetilde{h}(\bm{r})$ is a natural extension of the FP potential $V(Q)$ as a function of the global order parameter $Q$.

The stationary Equation~(\ref{results stationary}) reveals that the new closure (\ref{results closure}) corresponds to the mean-field equation of $W(\widetilde{G})$ given by Equation~(\ref{potential diff}).
The closure (\ref{results closure}) gives the self-consistent Equation~(\ref{scf}), similar to the previous one that predicts a dynamical transition~\cite{mor5,dfr4,dfr5};
we need to quantitatively assess the validity of Equation~(\ref{scf}) in terms of the dynamical transitions in simulation models.
Equation~(\ref{omega a1}) further suggests that we can go beyond the mean-field approximation as is the case with the Ginzburg-Landau-type theory:
the greatest advantage of our replica field theory is to systematically improve the self-consistent equation by considering fluctuations of inter-replica correlation field $\widetilde{h}(\bm{r})$.
It remains to be addressed whether the modified self-consistent equation explains the blurring of dynamical transition into a crossover from relaxational to activated~dynamics.

There is a caveat, turning our attention to the stability condition on $\widetilde{h}(\bm{r})$: translational and rotational symmetries are broken in frozen phases.
The violation becomes evident by expanding $\widetilde{h}(\bm{r})$ around that at the uniform density as follows~\cite{dft1,dft2,dft3}:
\begin{flalign}
\widetilde{h}(\bm{r}-\bm{r}';\rho^*(\bm{r}))
=\widetilde{h}(\bm{r}-\bm{r}';\overline{\rho})+\left.\int d\bm{r}'
\frac{\delta\widetilde{h}(\bm{r}-\bm{r}')}{\delta\rho}\right|_{\rho=\overline{\rho}}
\left\{\rho^*(\bm{r}')-\overline{\rho}\right\}+\cdots.
\label{h exp}
\end{flalign}
We also have a non-perturbative approach to avoid the difficulty using a globally weighted density $\overline{\rho}_{\rm{WD}}$ in the inter-replica TCF: $\widetilde{h}(\bm{r}-\bm{r}';\rho^*(\bm{r}))=\widetilde{h}(\bm{r}-\bm{r}';\overline{\rho}_{\rm{WD}})$, according to the modified weighted density functional approximation~\cite{dfg5,dfg10}.
Therefore, the~functional derivative in Equation~(\ref{h derivative}), or~the new closure (\ref{results closure}), holds approximately when either neglecting the second and higher-order terms in Equation~(\ref{h exp}) or finding $\overline{\rho}_{\rm{WD}}$.

The new closure (\ref{results closure}) in a metastable state provides the self-consistent Equation~(\ref{scf}) for the non-ergodicity parameter $f(\bm{k})$.
The present field theory has demonstrated the necessity to consider higher-order contributions in the perturbative treatment for obtaining the self-consistent Equation~(\ref{scf}) with a kernel containing the triplet DCF~\cite{dcf1,dcf2,dcf3}: we obtain Equation~(\ref{scf}) by adopting the approximate bridge function $B(\bm{r})=-\widetilde{h}^2(\bm{r})/2$ beyond the RHNC approximation of $B(\bm{r})=0$.
For comparison, we would like to mention two previous replica approaches to provide the triplet DCF in the self-consistent equation~\cite{mor5,dfr4,dfr5}.
The first approach considers the perturbative contribution to the replicated HNC functional along the liquid-state theory~\cite{mor5}.
The Legendre-transform-based method allows us to calculate the third order in $\widetilde{h}(\bm{r})$ concerning the Morita-Hiroike functional.
Meanwhile, the~second method considers the third-order term in density difference $\rho^*(\bm{r})-\overline{\rho}$ by taking the Ramakrishnan-Yussouf functional of the DFT as a reference form~\cite{dfr4,dfr5}.
Consequently, both perturbation methods amount to having the triplet DCF in the kernel of the self-consistent equation.
This agreement implies the equivalence between the replicated HNC and Ramakrishnan--Yussouf approximations, consistent with the conventional results of the liquid-state theory~\cite{ls1}.

Our scheme bears similarity to the Legendre-transform-based theory~\cite{mor1,mor2,mor3,mor4,mor5,mor6} rather than the DFT~\cite{dfr1,dfr2,dfr3,dfr4,dfr5}.
However, more elaborate input from the DFT~\cite{dft1,dft2,dft3} is also to be investigated, which is particularly necessary to investigate the glass transition in thin polymer films~\cite{film1,film2};
for example, we can improve the Ramakrishnan--Yussouf approximation by performing a variational evaluation beyond the Gaussian approximation (see Appendix \ref{gb}).
Furthermore, our replica theory has two additional features arising from the field-theoretical treatment of the inter-replica TCF $\widetilde{h}(\bm{r})$ and the associated two-body interaction potential $i\nu_a(\bm{r})$.
First, we can systematically consider fluctuations around the mean-field potential field, $\nu_a^*(\bm{r})=iu(\bm{r})$, given by Equation~(\ref{sp nu}), which is the same relation as that of the Legendre-transform-based method~\cite{mor1,mor2,mor3,mor4,mor5,mor6,morita,f4}.
Second, we can develop the replica field theory to include TCF fluctuations around the metastable field $\widetilde{h}_*(\bm{r})$ as described above.
Thus, the~present field-theoretical formalism opens up promising avenues to advance studies on the dynamical heterogeneity in terms of the correlation function of TCF fluctuations (i.e., the~so-called four-point correlation function~\cite{het1,het2,het3,het4}) as well as the FEL that includes fluctuations around a metastable state.

\newpage
\begin{center}
\large{\bf Abbreviations}
\end{center}
\begin{center}
\begin{tabular}{ll}
\hline
{\bf Acronym} & {\bf Definition} \\ 
\hline
FEL & free energy landscape \\
DFT & density functional theory \\
MCT & mode coupling theory \\
HNC approximation & hypernetted-chain approximation \\
RHNC & replicated hypernetted-chain \\
FP potential & Franz-Parisi potential\\
DCF & direct correlation function \\
TCF & total correlation function \\
rhs & right-hand side \\
MSA & mean spherical approximation \\
RPA & random phase approximation \\
\hline
\end{tabular}    
\end{center}

\newpage
\begin{center}
\large{\bf Supplemental Material}
\end{center}
\appendix
\section[\appendixname~\thesection]{Proof of {\itshape Result 1}: Derivation of Equation~(\ref{results replica}) Using the Replica Trick}\label{appendix replica}
Equation~(\ref{results omega2}) provides
\begin{flalign}
\frac{\beta \partial\Omega_{\bm{v}}^{(m)}}{\partial m}
&=-\frac{\partial}{\partial m}\left[\ln\left\{\mathrm{Tr}_1\,e^{-\beta U_{\mathrm{intra}}(\widehat{\rho}_1)-(m-1)\beta N^*W(\widetilde{G},\widehat{\rho}_1)}\right\}\right]
\nonumber\\
&=\frac{\mathrm{Tr}_1\,e^{-\beta U_{\mathrm{intra}}(\widehat{\rho}_1)}\beta N^*W(\widetilde{G},\widehat{\rho}_1)\,e^{-(m-1)\beta N^*W(\widetilde{G},\widehat{\rho}_1)}}{\mathrm{Tr}_1\,e^{-\beta U_{\mathrm{intra}}(\widehat{\rho}_1)-(m-1)\beta N^*W(\widetilde{G},\widehat{\rho}_1)}}
\label{app partial m}
\end{flalign}
Hence, we verify Equation~(\ref{results replica}) through the following derivation:
\begin{flalign}
\beta N^*W(\widetilde{G})
&=\lim_{m\rightarrow 1}\frac{\beta \partial\Omega_v^{(m)}}{\partial m}
\nonumber\\
&=\frac{\mathrm{Tr}_1\,e^{-\beta U_{\mathrm{intra}}(\widehat{\rho}_1)}\beta N^*W(\widetilde{G},\widehat{\rho}_1)}{\mathrm{Tr}_1\,e^{-\beta U_{\mathrm{intra}}(\widehat{\rho}_1)}}
\nonumber\\
&=\mathrm{Tr}_1\,P_{\rm{eq}}(\widehat{\rho}_1)\beta N^*W(\widetilde{G},\widehat{\rho}_1)
\nonumber\\
&=\beta N^*W(\widetilde{G}),
\label{app replica}
\end{flalign}
where the probability $P_{\rm{eq}}(\widehat{\rho}_1)$ is defined by Equation~(\ref{results peq}).
The last line of Equation~(\ref{app replica}) equals the rhs of the definition (\ref{def wg}).

\section[\appendixname~\thesection]{Strong-Coupling Perturbation Theory: General Formalism}\label{sc}
This appendix provides a general formalism of field theory for strongly-coupled glassy systems with the help of a variational~approach.
\subsection[\appendixname~\thesubsection]{Bare Interactions Mimicked by DCF: the Gibbs-Bogoliubov Inequality Approach}\label{gb}
We first consider the grand potential $e^{-\beta\Phi_{\bm{v}'}^{(m)}}=\boldsymbol{\mathrm{Tr}}\,e^{-\beta U(\bm{v}',\boldsymbol{\widehat{\rho}})}$ for the bare interaction potential matrix $\bm{v}'$ that is adjusted to satisfy Equation~(\ref{constraint2}) without the constraint $\Delta_a(\widetilde{G},\boldsymbol{\widehat{\rho}})$ by adding an inter-replica potential $\widetilde{v}(\bm{r})$ to the original potential matrix $\bm{v}$ where $\widetilde{v}(\bm{r})=0$.
We explore an optimized interaction potential to mimic the bare interactions represented by $\bm{v}'(\bm{r})$, using the Gibbs-Bogoliubov inequality for the lower bound~\cite{ls1,f1}:
\begin{flalign}
\beta\Phi_{\bm{v}'}^{(m)}\geq
\beta\Phi_{\bm{w}}^{(m)}
+\sum_{a=1}^m\sum_{b=1}^m\frac{1}{2}\int\!\!\!\int d\bm{r}d\bm{r}'
G_{ab}(\bm{r},\bm{r}')\left\{
\beta v_{ab}(\bm{r}-\bm{r}')-\beta w_{ab}(\bm{r}-\bm{r}')
\right\}.
\label{inequality}
\end{flalign}
Maximizing the rhs of Equation~(\ref{inequality}) with respect to $w_{ab}(\bm{r})$, we have
\begin{flalign}
\beta w_{ab}(\bm{r})=-c_{ab}(\bm{r}),
\label{wc}
\end{flalign}
in the Gaussian approximation, as~shown in Appendix \ref{appendix gaussian} using the density functional integral representation of $\Phi_{\bm{w}}^{(m)}$; see Equation~(\ref{app dft}) for the equivalence between various expressions of $\Phi_{\bm{w}}^{(m)}$.
Accordingly, the~grand potential $\Phi_{\bm{v}'}^{(m)}$ can be approximated by
\vspace{-6pt}
\begingroup\makeatletter\def\f@size{9}\check@mathfonts
\def\maketag@@@#1{\hbox{\m@th\normalsize\normalfont#1}}
\begin{flalign}
\beta\Phi_{\bm{v}'}^{(m)}
\approx
\beta\Phi_{-k_BT\bm{c}}^{(m)}
+\sum_{a=1}^m\sum_{b=1}^m\frac{1}{2}\int\!\!\!\int d\bm{r}d\bm{r}'
\rho_a^*(\bm{r})\rho_b^*(\bm{r}')g_{ab}(\bm{r}-\bm{r}')\left\{
\beta v_{ab}(\bm{r}-\bm{r}')+c_{ab}(\bm{r}-\bm{r}')
\right\}.
\label{maximum}
\end{flalign}
\endgroup
Equation~(\ref{maximum}) reads
\begin{flalign}
e^{-\beta\Phi_{\bm{v}'}^{(m)}}
=\boldsymbol{\mathrm{Tr}}\,e^{-\beta U(-k_BT\bm{c},\boldsymbol{\widehat{\rho}})
-\sum_{a=1}^m\sum_{b=1}^m\frac{1}{2}\int\!\!\!\int d\bm{r}d\bm{r}'
\rho_a^*(\bm{r})\rho_b^*(\bm{r}')g_{ab}(\bm{r}-\bm{r}')\left\{
\beta v_{ab}(\bm{r}-\bm{r}')+c_{ab}(\bm{r}-\bm{r}')
\right\}},
\label{phi dcf def}
\end{flalign}
using the original definition of  $e^{-\beta\Phi_{-k_BT\bm{c}}^{(m)}}=\boldsymbol{\mathrm{Tr}}\,e^{-\beta U(-k_BT\bm{c},\boldsymbol{\widehat{\rho}})}$.
For later convenience, we express $U(-k_BT\bm{c},\boldsymbol{\widehat{\rho}})$ as
\begin{flalign}
\label{def hc}
\beta U(-k_BT\bm{c},\boldsymbol{\widehat{\rho}})
&=\frac{1}{8}\left\{c_1({\bf 0})+(m-1)c({\bf 0})\right\}
\nonumber\\
&-\frac{1}{2}\int\!\!\!\int d\bm{r}d\bm{r}'\{
\boldsymbol{\widehat{\rho}}(\bm{r})-\boldsymbol{\widehat{\delta}}(\bm{r})
\}^{\rm{T}}
\bm{c}(\bm{r}-\bm{r}')
\left\{
\boldsymbol{\widehat{\rho}}(\bm{r}')-\boldsymbol{\widehat{\delta}}(\bm{r}')
\right\},
\end{flalign}
by introducing a self-energy operator,
\begin{flalign}
\label{def delta vec}
\boldsymbol{\widehat{\delta}}(\bm{r})=\frac{1}{2}
\left(
\begin{array}{c}
\delta_{1a}\delta_{\bm{0}}\,\widehat{\rho}_1^{(1)}(\bm{r})\\ 
\vdots \\
\delta_{ma}\delta_{\bm{0}}\,\widehat{\rho}_m^{(1)}(\bm{r})
\end{array}
\right),
\end{flalign}
where $\delta_{na}\delta_{\bm{0}}$ ($1\leq n\leq m$) represents the operator for an intra-replica interaction potential at zero separation.
The inhomogeneous Ornstein-Zernike equations given by Equations~(\ref{general oz})--(\ref{results oz inter}) are obtained from the set of maximum conditions on the variational interaction potential $w_{ab}(\bm{r})$ as follows:
\begin{flalign}
\label{second legendre}
\left.
\frac{\delta\Phi_{\bm{w}}^{(m)}}{\delta w_{ab}}
\right|_{w_{ab}=-k_BTc_{ab}}
=\frac{1}{2}\rho_a^*(\bm{r})\rho_b^*(\bm{r}')g_{ab}(\bm{r}),
\end{flalign}
similar to the relation used in the second Legendre transform (see Appendix \ref{appendix gaussian} for details).

\subsection[\appendixname~\thesubsection]{Evaluation Method of the Grand Potential Difference Due to the Constraint in Equation~(\ref{def omega})}\label{grand diff}

Combining Equations~(\ref{def omega}) and (\ref{phi dcf def}), the~grand potential difference $\Omega^{(m)}_{\bm{v}'}(\widetilde{G})-\Phi^{(m)}_{\bm{v}'}$ caused by the constraint $\prod_{a=2}^{m}\Delta_a(\widetilde{G},\boldsymbol{\widehat{\rho}})$ can be written as
\begin{flalign}
\label{delta omega approximation}
e^{-\beta\Omega^{(m)}_{\bm{v}'}(\widetilde{G})+\beta\Phi^{(m)}_{\bm{v}'}}
=\frac{\boldsymbol{\mathrm{Tr}}\,e^{-\beta U(-k_BT\bm{c},\boldsymbol{\widehat{\rho}})}\prod_{a=2}^{m}\Delta_a(\widetilde{G},\boldsymbol{\widehat{\rho}})}{\boldsymbol{\mathrm{Tr}}\,e^{-\beta U(-k_BT\bm{c},\boldsymbol{\widehat{\rho}})}}
\equiv
\left<
\prod_{a=2}^{m}\Delta_a(\widetilde{G},\boldsymbol{\widehat{\rho}})
\right>_{\bm{c}}.
\end{flalign}
Functional-integral representation of the constraint $\prod_{a=2}^{m}\Delta_a(\widetilde{G},\boldsymbol{\widehat{\rho}})$ in Equation~(\ref{def omega}) is
\begin{flalign}
&\prod_{a=2}^{m}\Delta_a(\widetilde{G},\boldsymbol{\widehat{\rho}})
\nonumber\\
&=\int D'\boldsymbol{\nu}
\exp\left[\,
\sum_{a=2}^m\int\!\!\!\int d\bm{r}d\bm{r}'
\,i \,\left\{
\rho_1^*(\bm{r})\rho^*(\bm{r}')\widetilde{g}(\bm{r}-\bm{r}')-\widehat{\rho}_1^{(N)}(\bm{r})\widehat{\rho}_a^{(N)}(\bm{r}')
\right\}\nu_a(\bm{r}-\bm{r}')\right],
\label{functional delta}
\end{flalign}
where $\prod_{a=2}^{m}\int D\nu_a\equiv\int D'\boldsymbol{\nu}$ and $\boldsymbol{\nu}=(\nu_2,\cdots,\nu_m)^{\mathrm{T}}$.
It follows from Equations~(\ref{delta omega approximation}) and (\ref{functional delta}) that
\begin{flalign}
e^{-\beta\Omega^{(m)}_{\bm{v}'}(\widetilde{G})+\beta\Phi^{(m)}_{\bm{v}'}}
=&\int D'\boldsymbol{\nu}\,
e^{\sum_{a=2}^m\int\!\!\!\int d\bm{r}d\bm{r}'
\,i \,\rho_1^*(\bm{r})\rho^*(\bm{r}')\widetilde{g}(\bm{r}-\bm{r}')\nu_a(\bm{r}-\bm{r}')}
\left<
e^{-\sum_{a=2}^m\int\!\!\!\int d\bm{r}d\bm{r}'\,i\,
\widehat{\rho}_1^{(N)}(\bm{r})\widehat{\rho}_a^{(N)}(\bm{r}')
\,\nu_a(\bm{r}-\bm{r}')
}
\right>_{\bm{c}}.
\label{delta omega approximation2}
\end{flalign}
We evaluate the average $\left<\mathcal{O}\right>_{\bm{c}}$ on the rhs of Equation~(\ref{delta omega approximation2}) by developing the strong-coupling perturbation~theory.

To this end, we introduce the auxiliary $\boldsymbol{\phi}$-field to have the functional-integral representation of $e^{-\beta U(-k_BT\bm{c},\boldsymbol{\widehat{\rho}})}$ as follows~\cite{text,fred1,fred2,matsen,euclid,f0,f2,f4,woo,russia}:
\begin{flalign}
e^{-\beta U(-k_BT\bm{c},\boldsymbol{\widehat{\rho}})}
&=\int D\boldsymbol{\rho}\,
e^{-\beta U(-k_BT\bm{c},\boldsymbol{\rho})}
\prod_{b=1}^m\prod_{\{\bm{r}\}}\delta\left[\rho_b(\bm{r})-\widehat{\rho}_b^{(N)}(\bm{r})
\right]
\nonumber\\
&=\iint D\boldsymbol{\phi}\,D\boldsymbol{\rho}\,
e^{-\beta U(-k_BT\bm{c},\boldsymbol{\rho})
+\int d\bm{r}\,i\,\boldsymbol{\phi}(\bm{r})\cdot
\left\{\boldsymbol{\widehat{\rho}}(\bm{r})-\boldsymbol{\rho}(\bm{r})\right\}}
\nonumber\\
&=\frac{1}{\mathcal{N}_c}\int D\boldsymbol{\phi}\,
e^{-\beta\mathcal{H}_0(\bm{c},\boldsymbol{\phi})
+\int d\bm{r}\,i\,\boldsymbol{\phi}(\bm{r})\cdot
\boldsymbol{\widehat{\rho}}(\bm{r})},
\label{potential int}
\end{flalign}
where $\boldsymbol{\phi}=(\phi_1,\cdots,\phi_m)^{\mathrm{T}}$, $\prod_{a=1}^m\int D\phi_a\equiv\int D\boldsymbol{\phi}$, the~Gaussian integration over the $\boldsymbol{\rho}$-field yields the normalization factor $\mathcal{N}_c$ written as
\begin{flalign}
\label{def norm c}
\mathcal{N}_c=\int D\boldsymbol{\phi}\,e^{\frac{1}{2}\int\!\!\!\int d\bm{r}d\bm{r}'
\boldsymbol{\phi}^{\rm{T}}(\bm{r})\bm{c}^{-1}(\bm{r}-\bm{r}')\boldsymbol{\phi}(\bm{r}')},
\end{flalign}
and
\begin{flalign}
\label{def h0}
\beta\mathcal{H}_0(\bm{c},\boldsymbol{\phi})&=
\frac{1}{8}\left\{c_1({\bf 0})+(m-1)c({\bf 0})\right\}
\nonumber\\
&-\frac{1}{2}\int\!\!\!\int d\bm{r}d\bm{r}'
\boldsymbol{\phi}^{\rm{T}}(\bm{r})\bm{c}^{-1}(\bm{r}-\bm{r}')\boldsymbol{\phi}(\bm{r}')
+\int d\bm{r}\,i\,\boldsymbol{\phi}(\bm{r})\cdot\boldsymbol{\widehat{\delta}}(\bm{r}).
\end{flalign}
Without particles (i.e., $\boldsymbol{\widehat{\rho}}=\bm{0}$), we have
\begin{flalign}
e^{-\beta U(-k_BT\bm{c},\boldsymbol{\widehat{\rho}}=0)}=
\frac{1}{\mathcal{N}_c}\int D\boldsymbol{\phi}\,
e^{-\beta\mathcal{H}_0(\bm{c},\boldsymbol{\phi})}=1,
\label{zero}
\end{flalign}
consistent with the trivial result $U(-k_BT\bm{c},\boldsymbol{\widehat{\rho}}=0)=0$ in Equation~(\ref{potential int}) (see also Appendix~\ref{appendix zero}).

Meanwhile, the~configurational integral represented by $\boldsymbol{\mathrm{Tr}}$ provides the perturbative contribution, $\mathcal{H}_1(\boldsymbol{\nu},\boldsymbol{\phi})$, given by
\begin{flalign}
\label{def h1}
e^{-\beta\mathcal{H}_1(\boldsymbol{\nu},\boldsymbol{\phi})}=
\boldsymbol{\mathrm{Tr}}\,
e^{\int d\bm{r}\,i \,\boldsymbol{\phi}(\bm{r})\cdot\boldsymbol{\widehat{\rho}}(\bm{r})
-\sum_{a=2}^m\int\!\!\!\int d\bm{r}d\bm{r}'\,i\,
\widehat{\rho}_1^{(N)}(\bm{r})\widehat{\rho}_a^{(N)}(\bm{r}')
\,\nu_a(\bm{r}-\bm{r}')}.
\end{flalign}
Defining the functional, 
\begin{flalign}
\label{def a nu}
e^{-\beta\mathcal{F}(\boldsymbol{\nu})}
=\frac{1}{\mathcal{N}_c}\int D\boldsymbol{\phi}\,
e^{-\beta\mathcal{H}_0(\bm{c},\boldsymbol{\phi})
-\beta\mathcal{H}_1(\boldsymbol{\nu},\boldsymbol{\phi})},
\end{flalign}
it follows from Equations~(\ref{delta omega approximation}) and (\ref{potential int}) that
\begin{flalign}
e^{-\beta\mathcal{F}(\boldsymbol{\nu}=\bm{0})}=e^{-\beta\Phi_{\bm{v'}}^{(m)}}=\boldsymbol{\mathrm{Tr}}\,e^{-\beta U(-k_BT\bm{c},\boldsymbol{\widehat{\rho}})}.
\label{f0 equality}
\end{flalign}
Combining Equations~(\ref{potential int})--(\ref{f0 equality}), we see that the average on the rhs of Equation~(\ref{delta omega approximation2}) simply reads
\begin{flalign}
\label{average nu}
e^{-\beta\mathcal{F}(\boldsymbol{\nu})+\beta\mathcal{F}(\boldsymbol{\nu}=\bm{0})}
=\left<
e^{-\sum_{a=2}^m\int\!\!\!\int d\bm{r}d\bm{r}'\,i\,
\widehat{\rho}_1^{(N)}(\bm{r})\widehat{\rho}_a^{(N)}(\bm{r}')
\,\nu_a(\bm{r}-\bm{r}')
}
\right>_{\bm{c}}.
\end{flalign}
Equations~(\ref{def h1})--(\ref{average nu}) indicate that the remaining task is to properly evaluate the perturbative contribution $\mathcal{H}_1(\boldsymbol{\nu},\boldsymbol{\phi})$ in Equation~(\ref{def a nu}) at strong~coupling.

\subsection[\appendixname~\thesubsection]{Strong-Coupling Expansion}\label{sc section}
Equation~(\ref{results mean}) implies that
\begin{flalign}
\label{fugacity1}
e^{\beta\mu-\frac{c_{aa}({\bf 0})}{2}}
=\frac{N^*}{
\int d\bm{r}e^{
\sum_{b=1}^m\int d\bm{r}'c_{ab}(\bm{r}-\bm{r}')\rho_b^*(\bm{r}')}
}
\approx \overline{\rho},
\end{flalign}
considering the finite range of the DCF.
It follows from Equation~(\ref{fugacity1}) that the fugacity $e^{\beta\mu}$ becomes
\begin{flalign}
\label{fugacity2}
e^{\beta\mu}\approx\frac{\overline{\rho}}{\gamma_a}
\end{flalign}
by introducing the coupling parameter $\gamma_a$ defined as follows:
\begin{flalign}
\label{def gamma}
\gamma_a=e^{-\frac{c_{aa}({\bf 0})}{2}},
\end{flalign}
or $\gamma_1=e^{-\frac{c_{1}({\bf 0})}{2}}$ for $a=1$ and $\gamma=e^{-\frac{c({\bf 0})}{2}}$ for $a\geq 2$.

The relation (\ref{fugacity2}) clarifies that the fugacity expansion method is validated at strong coupling, or~in the glassy state. 
In what follows, we set $\gamma=\gamma_a$ for brevity considering that the glassy state is located in the strong-coupling regime of $\gamma_1,\,\gamma\gg 1$ because of $-c_1(\bm{0}),\,-c(\bm{0})\gg 1$ at freezing~\cite{f3}.
The strong-coupling perturbation theory is based on the use of the following expansion:
\vspace{-6pt}
\begingroup\makeatletter\def\f@size{9}\check@mathfonts
\def\maketag@@@#1{\hbox{\m@th\normalsize\normalfont#1}}
\begin{flalign}
\boldsymbol{\mathrm{Tr}}=1+\frac{\overline{\rho}}{\gamma}\sum_{a=1}^m\int d\bm{r}_{a,1}
+\left(
\frac{\overline{\rho}}{\gamma}
\right)^2\left(
\frac{1}{2}\sum_{a=1}^m\int\!\!\!\int d\bm{r}_{a,1}d\bm{r}_{a,2}
+\sum_{a(>b)}\sum_{b=1}^m\int\!\!\!\int d\bm{r}_{a,1}d\bm{r}_{b,1}
\right)+\mathcal{O}\left[\gamma^{-3}\right].
\label{tr exp}
\end{flalign}
\endgroup
Substituting Equation~(\ref{tr exp}) into Equation~(\ref{def h1}), we obtain
\begin{flalign}
-\beta\mathcal{H}_1(\boldsymbol{\nu},\boldsymbol{\phi})
&=\ln\left\{
\boldsymbol{\mathrm{Tr}}\,
e^{\int d\bm{r}\,i \,\boldsymbol{\phi}(\bm{r})\cdot\boldsymbol{\widehat{\rho}}(\bm{r})
-\sum_{a=2}^m\int\!\!\!\int d\bm{r}d\bm{r}'\,i\,
\widehat{\rho}_a^{(N)}(\bm{r})\widehat{\rho}_1^{(N)}(\bm{r}')
\,\nu_a(\bm{r}-\bm{r}')}
\right\}\nonumber\\
&\approx\ln\left\{1+\frac{\overline{\rho}}{\gamma}\mathcal{U}_1(\boldsymbol{\phi})
+\left(\frac{\overline{\rho}}{\gamma}\right)^2
\mathcal{U}_2(\boldsymbol{\phi})
\right\}\nonumber\\
&\approx\frac{\overline{\rho}}{\gamma}\mathcal{U}_1(\boldsymbol{\phi})
+\left(\frac{\overline{\rho}}{\gamma}\right)^2\left\{
\mathcal{U}_2(\boldsymbol{\phi})-\frac{1}{2}\mathcal{U}_1^2(\boldsymbol{\phi})\right\},
\label{gamma exp}
\end{flalign}
where
\begin{flalign}
\label{def u1}
\mathcal{U}_1(\boldsymbol{\phi})
&=\sum_{a=1}^m\int d\bm{r}_{a,1}\,e^{\int d\bm{r}
i \phi_{a}(\bm{r})\widehat{\rho}_a^{(1)}(\bm{r})},
\\
\mathcal{U}_2(\boldsymbol{\phi})
&=\frac{1}{2}\sum_{a=1}^m\int\!\!\!\int d\bm{r}_{a,1}d\bm{r}_{a,2}
\,e^{\int d\bm{r}
i \phi_{a}(\bm{r})\widehat{\rho}_a^{(2)}(\bm{r})
}\nonumber\\
&+\sum_{a(>b)}\sum_{b=2}^m\int\!\!\!\int d\bm{r}_{a,1}d\bm{r}_{b,1}
\,e^{\int d\bm{r}\,i \phi_{a}(\bm{r})
\left\{\widehat{\rho}_a^{(1)}(\bm{r})+\widehat{\rho}_b^{(1)}(\bm{r})\right\}}
\nonumber\\
\label{def u2}
&+\sum_{a=2}^m\int\!\!\!\int d\bm{r}_{a,1}d\bm{r}_{1,1}
\,e^{\int d\bm{r}\,
\left\{i \phi_1(\bm{r})\widehat{\rho}_1^{(1)}(\bm{r})
+i \phi_a(\bm{r})\widehat{\rho}_a^{(1)}(\bm{r})\right\}
-\int\!\!\!\int d\bm{r}d\bm{r}'\,i\,\widehat{\rho}_1^{(1)}(\bm{r})\widehat{\rho}_a^{(1)}(\bm{r}')\nu_a(\bm{r}-\bm{r}')
}.
\end{flalign}
Rearranging the terms in Equations~(\ref{def u1}) and (\ref{def u2}), we have
\begin{flalign}
\mathcal{U}_2(\boldsymbol{\phi})-\frac{1}{2}\mathcal{U}_1^2(\boldsymbol{\phi})
=\sum_{a=2}^m\int\!\!\!\int d\bm{r}_{1,1}d\bm{r}_{a,1}
\,e^{\int d\bm{r}\,
\left\{i \phi_1(\bm{r})\widehat{\rho}_1^{(1)}(\bm{r})
+i \phi_a(\bm{r})\widehat{\rho}_a^{(1)}(\bm{r})\right\}}
f(i\nu_a),
\label{u1 u2 reduction}
\end{flalign}
using the Mayer $f$-function:
\begin{flalign}
f(i\nu_a)=e^{-i\nu_a(\bm{r}_{1,1}-\bm{r}_{a,1})}-1.
\label{def mayer f}
\end{flalign}
It is found from Equations~(\ref{gamma exp}) and (\ref{u1 u2 reduction}) that Equation~(\ref{def a nu}) is approximated by
\begin{flalign}
e^{-\beta\mathcal{F}(\boldsymbol{\nu})}
=\frac{1}{\mathcal{N}_c}\int D\boldsymbol{\phi}\,
e^{-\beta\mathcal{H}_{\mathrm{mf}}(\boldsymbol{\phi})}
\left\{
1+\left(\frac{\overline{\rho}}{\gamma}\right)^2
\sum_{a=2}^m\int\!\!\!\int d\bm{r}d\bm{r}'
\,e^{i \phi_1(\bm{r})+i \phi_a(\bm{r}')}
f(i\nu_a)
\right\},
\label{sc expansion}
\end{flalign}
where
\begin{flalign}
\beta\mathcal{H}_{\mathrm{mf}}(\boldsymbol{\phi})=
\beta\mathcal{H}_0(\bm{c},\boldsymbol{\phi})
-\frac{\overline{\rho}}{\gamma}\mathcal{U}_1(\boldsymbol{\phi}).
\label{def a}
\end{flalign}
Equations~(\ref{f0 equality}), (\ref{def mayer f}) and (\ref{sc expansion}) imply that
\begin{flalign}
e^{-\beta\Phi_{-k_BT\bm{c}}^{(m)}}=e^{-\beta\mathcal{F}(\boldsymbol{\nu}=\bm{0})}
=\frac{1}{\mathcal{N}_c}\int D\boldsymbol{\phi}\,
e^{-\beta\mathcal{H}_{\mathrm{mf}}(\boldsymbol{\phi})}
\label{f0 integral}
\end{flalign}
because of $f(i\nu_a=0)=0$.
Combining Equations~(\ref{sc expansion}) and (\ref{f0 integral}), we obtain the following approximate result at strong coupling:
\begin{flalign}
\beta\mathcal{F}(\boldsymbol{\nu})-\beta\mathcal{F}(\boldsymbol{\nu}=\bm{0})
=-\left(\frac{\overline{\rho}}{\gamma}\right)^2
\sum_{a=2}^m\int\!\!\!\int d\bm{r}_{1,1}d\bm{r}_{a,1}
\,\left<
e^{\int d\bm{r}\,
\left\{i \phi_1(\bm{r})\widehat{\rho}_1^{(1)}(\bm{r})
+i \phi_a(\bm{r})\widehat{\rho}_a^{(1)}(\bm{r})\right\}}
\right>_{\phi}
f(i\nu_a),
\label{sc main}
\end{flalign}
where the subscript $\phi$ denotes the averaging procedure as follows:
\begin{flalign}
\left<\mathcal{O}\right>_{\phi}
=\frac{\int D\boldsymbol{\phi}\,\mathcal{O}\,
e^{-\beta\mathcal{H}_{\mathrm{mf}}(\boldsymbol{\phi})}}
{\int D\boldsymbol{\phi}\,
e^{-\beta\mathcal{H}_{\mathrm{mf}}(\boldsymbol{\phi})}}.
\label{phi average}
\end{flalign}
Equation~(\ref{sc main}) is a representative result of the strong-coupling perturbation theory developed in this~section.

\subsection[\appendixname~\thesubsection]{Verifying Equation~(\ref{wc}) in the Gaussian Approximation}\label{appendix gaussian}
To evaluate density-density correlations, it is straightforward to use the density functional integral representation of $\Phi^{(m)}_{\bm{w}}(\widetilde{G})$ expressed as~\cite{f0,f2,f4,woo,russia}
\begin{flalign}
e^{-\beta \Phi_{\bm{w}}^{(m)}(\widetilde{G})}
&=\boldsymbol{\mathrm{Tr}}\,
e^{-\beta U(\bm{w},\boldsymbol{\widehat{\rho}})}
\nonumber\\
&=\int D\boldsymbol{\rho}\, \boldsymbol{\mathrm{Tr}}\,
e^{-\beta U(\bm{w},\boldsymbol{\rho})}
\prod_{b=1}^m\prod_{\{\bm{r}\}}\delta\left[\rho_b(\bm{r})-\widehat{\rho}_b^{(N)}(\bm{r})
\right]
\nonumber\\
&=\int D\boldsymbol{\rho}\,
e^{-\beta F_{\rm{mf}}(\bm{w},\boldsymbol{\rho})},
\label{app dft}
\end{flalign}
using the density functional $F_{\rm{mf}}(\bm{w},\boldsymbol{\rho})$ defined by Equation~(\ref{def mf}).
Equation~(\ref{app dft}) leads to
\begin{flalign}
\frac{\delta\Phi_{\bm{w}}^{(m)}}{\delta w_{ab}}=\frac{1}{2}\left<\rho_a(\bm{r})\rho_b(\bm{r}')\right>_{\rho}
\label{app second legendre}
\end{flalign}
for the left-hand side of Equation~(\ref{second legendre}).
The subscript $\rho$ in Equation~(\ref{app second legendre}) represents the following average:
\begin{flalign}
\left<\mathcal{O}\right>_{\rho}=\frac{\int D\boldsymbol{\rho}\,\mathcal{O}\,
e^{-\beta F_{\rm{mf}}(\bm{w},\boldsymbol{\rho})}}{e^{-\beta \Phi_{\bm{w}}^{(m)}(\widetilde{G})}}.
\label{app dmean}
\end{flalign}
It is also noted that the saddle-point equation $\delta F_{\rm{mf}}/\delta\rho_a|_{\rho=\rho_a^*}=0$ in Equation~(\ref{app dft}) provides
\begin{flalign}
\rho_a^*(\bm{r})=e^{\beta\mu+\frac{\beta w_{aa}({\bf 0})}{2}}
\exp\left\{
-\sum_{b=1}^m\int d\bm{r}'\beta w_{ab}(\bm{r}-\bm{r}')\rho_b^*(\bm{r}')
\right\},
\label{app mf density}
\end{flalign}
which matches Equation~(\ref{results mean}) when Equation~(\ref{wc}) is~verified.

Let $\bm{n}=\boldsymbol{\rho}-\boldsymbol{\rho}^*$ be a fluctuating density vector.
In the saddle-point approximation, the~mean density-density correlation appearing in the rhs of Equation~(\ref{app second legendre}) reads
\begin{flalign}
\left<\rho_a(\bm{r})\rho_b(\bm{r}')\right>_{\rho}=\rho_a^*(\bm{r})\rho_b^*(\bm{r}')+\left<n_a(\bm{r})n_b(\bm{r}')\right>_{\rho}.
\label{app element}
\end{flalign}
The correlation function $N_{ab}(\bm{r}-\bm{r}')=\left<n_a(\bm{r})n_b(\bm{r}')\right>_{\rho}$ in Equation~(\ref{app element}) corresponds to the matrix element of density-density correlation matrix $\bm{N}=\left<\bm{n}\bm{n}^{\rm{T}}\right>_{\rho}$.
If the equality (\ref{second legendre}) holds, Equations~(\ref{app second legendre}) and (\ref{app element}) imply that $N_{ab}(\bm{r}-\bm{r}')$ is related to the TCF $h_{ab}(\bm{r})=g_{ab}(\bm{r})-1$ as
\begin{flalign}
N_{ab}(\bm{r}-\bm{r}')=
\rho_a^*(\bm{r})\delta_{ab}\delta(\bm{r}-\bm{r}')
+\rho_a^*(\bm{r})\rho_b^*(\bm{r}')h_{ab}(\bm{r}-\bm{r}').
\label{app nab}
\end{flalign}
Meanwhile, the~matrix element $N_{ab}^{-1}(\bm{r}-\bm{r}')$ of the inverse matrix $\bm{N}^{-1}$ is given by
\begin{flalign}
N_{ab}^{-1}(\bm{r}-\bm{r}')=\beta w_{ab}(\bm{r}-\bm{r}')+\frac{\delta_{ab}\delta(\bm{r}-\bm{r}')}{\rho_a^*(\bm{r})},
\label{app n-1}
\end{flalign}
when considering Gaussian fluctuations of $\bm{n}(\bm{r})$.

The inverse matrix $\bm{N}^{-1}(\bm{r}-\bm{r}")$ satisfies $\int d\bm{r}"\bm{N}^{-1}(\bm{r}-\bm{r}")\bm{N}(\bm{r}"-\bm{r}')=\delta(\bm{r}-\bm{r}')\bm{I}$, or~\begin{flalign}
\label{app inverse}
\sum_{b=1}^m\int d\bm{r}"N_{ab}^{-1}(\bm{r}-\bm{r}")N_{bc}(\bm{r}"-\bm{r}')=\delta_{ac}\delta(\bm{r}-\bm{r}').
\end{flalign}
Substituting Equations~(\ref{app nab}) and (\ref{app n-1}) into Equation~(\ref{app inverse}), the~left-hand side of Equation~(\ref{app inverse}) is reduced to
\begin{flalign}
&\sum_{b=1}^m\int d\bm{r}"
\left\{\beta w_{ab}(\bm{r}-\bm{r}")
+\frac{\delta_{ab}\delta(\bm{r}-\bm{r}")}{\rho_a^*(\bm{r})}\right\}
\rho_b^*(\bm{r}")\delta_{bc}\delta(\bm{r}"-\bm{r}')\nonumber\\
&\qquad+\sum_{b=1}^m\int d\bm{r}"
\left\{\beta w_{ab}(\bm{r}-\bm{r}")
+\frac{\delta_{ab}\delta(\bm{r}-\bm{r}")}{\rho_a^*(\bm{r})}\right\}
\rho_b^*(\bm{r}")\rho_c^*(\bm{r}')h_{bc}(\bm{r}"-\bm{r}')\nonumber\\
&=\rho_c^*(\bm{r}')\beta w_{ac}(\bm{r}-\bm{r}')+\delta_{ac}\delta(\bm{r}-\bm{r}')+\sum_{b=1}^m\int d\bm{r}"\rho_b^*(\bm{r}")\rho_c^*(\bm{r}')\beta w_{ab}(\bm{r}-\bm{r}")h_{bc}(\bm{r}"-\bm{r}')
+\rho_c^*(\bm{r}')h_{ac}(\bm{r}-\bm{r}').
\label{app inverse2}
\end{flalign}
It follows from Equations~(\ref{app inverse}) and (\ref{app inverse2}) that
\begin{flalign}
h_{ac}(\bm{r}-\bm{r}')
=-\beta w_{ac}(\bm{r}-\bm{r}')
-\sum_{b=1}^m\int d\bm{r}"\rho_b^*(\bm{r}")\beta w_{ab}(\bm{r}-\bm{r}")h_{bc}(\bm{r}"-\bm{r}'),
\label{app inverse3}
\end{flalign}
which becomes the inhomogeneous Ornstein-Zernike Equation~(\ref{general oz}) when Equation~(\ref{wc}) holds.
Namely, combination of the maximum condition (\ref{second legendre}) with the inhomogeneous Ornstein-Zernike Equation~(\ref{general oz}) verifies Equation~(\ref{wc}).

\subsection[\appendixname~\thesubsection]{Derivation of Equation~(\ref{zero})}\label{appendix zero}
Let us rearrange $\mathcal{H}_0(\bm{c},\boldsymbol{\phi})$ in Equation~(\ref{def h0}) by completing the square. To~this end, we define an imaginary potential vector,
\begin{flalign}
\boldsymbol{\zeta}_c(\bm{r})=i\int d\bm{r}'\bm{c}(\bm{r}-\bm{r}')\boldsymbol{\widehat{\delta}}(\bm{r}'),
\label{appendix zetac}
\end{flalign}
with which Equation~(\ref{def h0}) is transformed into
\begin{flalign}
\beta\mathcal{H}_0(\bm{c},\boldsymbol{\phi})=
&-\frac{1}{2}\int\!\!\!\int d\bm{r}d\bm{r}'
\left\{\boldsymbol{\phi}^{\rm{T}}(\bm{r})
-\boldsymbol{\zeta}_c^{\rm{T}}(\bm{r})\right\}
\bm{c}^{-1}(\bm{r}-\bm{r}')\left\{\boldsymbol{\phi}(\bm{r}')
-\boldsymbol{\zeta}_c(\bm{r}')\right\},
\label{appendix square1}
\end{flalign}
noting that
\begin{flalign}
-\frac{1}{2}\int\!\!\!\int d\bm{r}d\bm{r}'
\boldsymbol{\zeta}_c^{\rm{T}}(\bm{r})
\bm{c}^{-1}(\bm{r}-\bm{r}')\boldsymbol{\zeta}_c(\bm{r}')
=\frac{1}{8}\left\{c_1({\bf 0})+(m-1)c({\bf 0})\right\}.
\label{appendix energy zero}
\end{flalign}
It is found from Equation~(\ref{appendix square1}) that 
\begin{flalign}
\int D\boldsymbol{\phi}\,
e^{-\beta\mathcal{H}_0(\bm{c},\boldsymbol{\phi})}=\mathcal{N}_c,
\label{appendix nc}
\end{flalign}
which is equivalent to Equation~(\ref{zero}).

\section[\appendixname~\thesection]{Derivation of Equation~(\ref{saddle total})}\label{appendix mf}
The key functional $\mathcal{H}_{\mathrm{mf}}(\boldsymbol{\phi})$ in the $\boldsymbol{\phi}$--functional integral of Equation~(\ref{sc expansion}) consists of two terms as seen from Equation~(\ref{def a}).
The first contribution $\beta\mathcal{H}_0(\bm{c},\boldsymbol{\phi})$ to $\mathcal{H}_{\mathrm{mf}}(\boldsymbol{\phi})$ at the saddle-point path $\boldsymbol{\phi}=i\boldsymbol{\psi}^*$ is written as
\begin{flalign}
&\beta\mathcal{H}_0(\bm{c},i\boldsymbol{\psi}^*)-\frac{1}{8}\left\{c_1({\bf 0})+(m-1)c({\bf 0})\right\}
\nonumber\\
&\qquad=\frac{1}{2}
\int\!\!\!\int d\bm{r}d\bm{r}'\boldsymbol{\psi}^*(\bm{r})^{\rm{T}}\bm{c}^{-1}(\bm{r}-\bm{r}')
\boldsymbol{\psi}^*(\bm{r}')
-\int d\bm{r}\,
\boldsymbol{\psi}^*(\bm{r})\cdot\boldsymbol{\widehat{\delta}}(\bm{r})
\nonumber\\
&\qquad=\frac{1}{2}
\int\!\!\!\int d\bm{r}d\bm{r}'\boldsymbol{\rho}^*(\bm{r})^{\rm{T}}\bm{c}(\bm{r}-\bm{r}')
\boldsymbol{\rho}^*(\bm{r}')
-\int d\bm{r}\,
\boldsymbol{\psi}^*(\bm{r})\cdot\boldsymbol{\widehat{\delta}}(\bm{r})
\nonumber\\
&\qquad\qquad\qquad\qquad\qquad\qquad\qquad
-\sum_{a=1}^m\int d\bm{r}\,\frac{\rho_a^*(\bm{r})}{2}c({\bf 0})
+\frac{1}{8}\left\{c_1({\bf 0})+(m-1)c({\bf 0})\right\}
\nonumber\\
&\qquad=\frac{1}{2}
\int\!\!\!\int d\bm{r}d\bm{r}'\boldsymbol{\rho}^*(\bm{r})^{\rm{T}}\bm{c}(\bm{r}-\bm{r}')
\boldsymbol{\rho}^*(\bm{r}')
-\frac{1}{4}\left\{c_1({\bf 0})+(m-1)c({\bf 0})\right\}
+\sum_{a=1}^m\int d\bm{r}\,\frac{\rho_a^*(\bm{r})}{2}c({\bf 0})
\nonumber\\
&\qquad\qquad\qquad\qquad\qquad\qquad\qquad
-\sum_{a=1}^m\int d\bm{r}\,\frac{\rho_a^*(\bm{r})}{2}c({\bf 0})
+\frac{1}{8}\left\{c_1({\bf 0})+(m-1)c({\bf 0})\right\}
\nonumber\\
&\qquad=\frac{1}{2}
\int\!\!\!\int d\bm{r}d\bm{r}'\boldsymbol{\rho}^*(\bm{r})^{\rm{T}}\bm{c}(\bm{r}-\bm{r}')
\boldsymbol{\rho}^*(\bm{r}')-\frac{1}{8}\left\{c_1({\bf 0})+(m-1)c({\bf 0})\right\},
\label{appendix saddle h0}
\end{flalign}
where use has been made of the relation, $\rho_a^*(\bm{r})=e^{\beta\mu-\psi_a^*(\bm{r})}$.
Hence, Equation~(\ref{def a}) leads to
\begin{flalign}
\mathcal{H}_{\mathrm{mf}}(i\boldsymbol{\psi}^*)
&=\beta\mathcal{H}_0(\bm{c},i\boldsymbol{\psi}^*)
-\frac{\overline{\rho}}{\gamma}\mathcal{U}_1(i\boldsymbol{\psi}^*)
\nonumber\\
&=\frac{1}{2}
\int\!\!\!\int d\bm{r}d\bm{r}'\boldsymbol{\rho}^*(\bm{r})^{\rm{T}}\bm{c}(\bm{r}-\bm{r}')
\boldsymbol{\rho}^*(\bm{r}')
-\sum_{a=1}^m\int d\bm{r}\,\rho_a^*(\bm{r})\nonumber\\
&=\beta F_{\rm{mf}}(-k_BT\bm{c},\boldsymbol{\rho}^*).
\label{appendix saddle total}
\end{flalign}
Thus, we verify the equality in Equation~(\ref{saddle total}) from Equations~(\ref{appendix saddle h0}) and (\ref{appendix saddle total}) in~detail.

\section[\appendixname~\thesection]{Derivation of Equation~(\ref{h-1c-1})}\label{appendix inverse oz}
We can demonstrate that the inhomogeneous Ornstein-Zernike Equation~(\ref{general oz}) is equivalent to Equation~(\ref{h-1c-1}).
First, Equation~(\ref{general oz}) becomes
\vspace{-6pt}
\begingroup\makeatletter\def\f@size{9}\check@mathfonts
\def\maketag@@@#1{\hbox{\m@th\normalsize\normalfont#1}}
\begin{flalign}
&\sum_{b=1}^m\int d\bm{r}"\,c_{ab}^{-1}(\bm{r}-\bm{r}")h_{bc}(\bm{r}"-\bm{r}')
\nonumber\\
&=\sum_{b=1}^m\int d\bm{r}"\left\{
\,c_{ab}^{-1}(\bm{r}-\bm{r}")c_{bc}(\bm{r}"-\bm{r}')
+\sum_{d=1}^m\int d\bm{u}\,
\rho^*_d(\bm{u})c_{ab}^{-1}(\bm{r}-\bm{r}")c_{bd}(\bm{r}"-\bm{u}))h_{dc}(\bm{u}-\bm{r}')
\right\}\nonumber\\
&=\delta_{ac}\delta(\bm{r}-\bm{r}')
+\sum_{d=1}^m\int d\bm{u}\,
\rho^*_d(\bm{u})\delta_{ad}\delta(\bm{r}-\bm{u}))h_{dc}(\bm{u}-\bm{r}')
\nonumber\\
&=\delta_{ac}\delta(\bm{r}-\bm{r}')
+\rho^*_a(\bm{r})h_{ac}(\bm{r}-\bm{r}').
\label{appendix oz transform1}
\end{flalign}
\endgroup
Furthermore, both the left-hand side and the rhs in the last line of Equation~(\ref{appendix oz transform1}) are transformed into
\begin{flalign}
&\sum_{b=1}^m\sum_{c=1}^m\int\!\!\!\int 
d\bm{r}"d\bm{r}'\,c_{ab}^{-1}(\bm{r}-\bm{r}")h_{bc}(\bm{r}"-\bm{r}')h_{cd}^{-1}(\bm{r}'-\bm{u})
\nonumber\\\
&\qquad=\sum_{b=1}^m\int d\bm{r}"\,
c_{ab}^{-1}(\bm{r}-\bm{r}")\delta_{bd}\delta(\bm{r}"-\bm{u})
\nonumber\\
&\qquad=c_{ad}^{-1}(\bm{r}-\bm{u})
\label{appendix oz transform2}
\end{flalign}
and
\begin{flalign}
\sum_{c=1}^m\int d\bm{r}'\,\left\{
\delta_{ac}\delta(\bm{r}-\bm{r}')h_{cd}^{-1}(\bm{r}'-\bm{u})
+\rho^*_a(\bm{r})h_{ac}(\bm{r}-\bm{r}')h_{cd}^{-1}(\bm{r}'-\bm{u})
\right\}
\nonumber\\
=h_{ad}^{-1}(\bm{r}-\bm{u})
+\rho^*_a(\bm{r})\delta_{ad}\delta(\bm{r}-\bm{u}),
\label{appendix oz transform3}
\end{flalign}
respectively.
Thus, we have proved that the inhomogeneous Ornstein-Zernike Equation~(\ref{general oz}) reads Equation~(\ref{h-1c-1}).

\section[\appendixname~\thesection]{Derivation of Equation~(\ref{sc sp})}\label{appendix sc}
We consider $m$ replicas that have two particles in total:
there is one particle, respectively, in~replica 1 and replica $a$, and~no particle exists in the other replicas.
Let $\boldsymbol{\widehat{\rho}}_{(1)}(\bm{r})\equiv(\rho_{1,1}^{(1)}(\bm{r}),0,\cdots,0,\rho_{a,1}^{(1)}(\bm{r}),0,\cdots)$ be the one-particle density vector in this system.
We need to rearrange the sum of $\beta\mathcal{H}_0(\bm{h},\boldsymbol{\varphi})$ and a one-particle energy $-i\int d\bm{r}\,\boldsymbol{\varphi}(\bm{r})\cdot\boldsymbol{\widehat{\rho}}_{(1)}(\bm{r})$ for the $\varphi$--averaging in Equation~(\ref{sc sp}).
For later convenience, we introduce a fluctuating potential that shifts to
\begin{flalign}
\boldsymbol{\varphi}_{\zeta}(\bm{r})=\boldsymbol{\varphi}(\bm{r})+\boldsymbol{\zeta}_1(\bm{r})-\boldsymbol{\zeta}_h(\bm{r})
\label{appendix shift}
\end{flalign}
using a reference potential,
\begin{flalign}
\boldsymbol{\zeta}_1(\bm{r})=i\int d\bm{r}'\bm{h}(\bm{r}-\bm{r}')\boldsymbol{\widehat{\rho}}_{(1)}(\bm{r}'),
\label{appendix one potential}
\end{flalign}
in addition to
\begin{flalign}
\boldsymbol{\zeta}_h(\bm{r})=i\int d\bm{r}'\bm{h}(\bm{r}-\bm{r}')\boldsymbol{\widehat{\delta}}(\bm{r}'),
\label{appendix zetah}
\end{flalign}
similar to Equation~(\ref{appendix zetac}).

We note that
\begin{flalign}
-\int\!\!\!\int d\bm{r}d\bm{r}'
\boldsymbol{\zeta}_h^{\rm{T}}(\bm{r})
\bm{h}^{-1}(\bm{r}-\bm{r}')\boldsymbol{\zeta}_1(\bm{r}')
=\frac{1}{2}\int d\bm{r}
\boldsymbol{\widehat{\rho}}_{(1)}^{\rm{T}}(\bm{r})
\bm{h}(\bm{r}-\bm{r}')\delta(\bm{r}-\bm{r}')
\label{appendix energy zero2}
\end{flalign}
and that
\begin{flalign}
-\frac{1}{2}\int\!\!\!\int d\bm{r}d\bm{r}'
\boldsymbol{\zeta}_h^{\rm{T}}(\bm{r})
\bm{h}^{-1}(\bm{r}-\bm{r}')\boldsymbol{\zeta}_h(\bm{r}')
=\frac{1}{8}\left\{h_1({\bf 0})+(m-1)h({\bf 0})\right\},
\label{appendix energy zero3}
\end{flalign}
similar to Equation~(\ref{appendix energy zero}).
Completing the square, we have
\begin{flalign}
\beta\mathcal{H}_0(\bm{h},\boldsymbol{\varphi})-i\int d\bm{r}\,\boldsymbol{\varphi}(\bm{r})\cdot\boldsymbol{\widehat{\rho}}^{(1)}(\bm{r})
=-\frac{1}{2}\int\!\!\!\int d\bm{r}d\bm{r}'
\boldsymbol{\varphi}_{\zeta}^{\rm{T}}(\bm{r})
\bm{h}^{-1}(\bm{r}-\bm{r}')\boldsymbol{\varphi}_{\zeta}(\bm{r}')
+\beta\mathcal{H}_{\zeta},
\label{appendix square2}
\end{flalign}
where
\begin{flalign}
\beta\mathcal{H}_{\zeta}
=&\frac{1}{2}\int\!\!\!\int d\bm{r}d\bm{r}'\left\{\boldsymbol{\zeta}_1(\bm{r})-\boldsymbol{\zeta}_h(\bm{r})\right\}^{\rm{T}}\bm{h}^{-1}(\bm{r}-\bm{r}')\left\{\boldsymbol{\zeta}_1(\bm{r}')-\boldsymbol{\zeta}_h(\bm{r}')\right\}
\nonumber\\
&
-\frac{1}{2}\int\!\!\!\int d\bm{r}d\bm{r}'
\boldsymbol{\zeta}_h^{\rm{T}}(\bm{r})
\bm{h}^{-1}(\bm{r}-\bm{r}')\boldsymbol{\zeta}_h(\bm{r}')
\nonumber\\
=&-\frac{1}{2}\int\!\!\!\int d\bm{r}d\bm{r}'\left\{
\boldsymbol{\widehat{\rho}}_{(1)}^{\rm{T}}(\bm{r})
\bm{h}(\bm{r}-\bm{r}')\boldsymbol{\widehat{\rho}}_{(1)}(\bm{r}')
-\boldsymbol{\widehat{\rho}}_{(1)}^{\rm{T}}(\bm{r})
\bm{h}(\bm{r}-\bm{r}')\delta(\bm{r}-\bm{r}')
\right\}
\nonumber\\
=&-\int\!\!\!\int d\bm{r}d\bm{r}'\widehat{\rho}_{1,1}^{(1)}(\bm{r})\widehat{\rho}_{a,1}^{(1)}(\bm{r})
\widetilde{h}(\bm{r}-\bm{r}')=-\widetilde{h}(\bm{r}_{1,1}-\bm{r}_{a,1}),
\label{appendix hzeta}
\end{flalign}
thus verifying the result in Equation~(\ref{sc sp}).

\section[\appendixname~\thesection]{Verifying the First Term on the rhs of Equation~(\ref{h derivative})}\label{appendix ln}
The Ornstein-Zernike Equation~(\ref{results oz inter}) at $m=1$ yields approximately
\vspace{-6pt}
\begingroup\makeatletter\def\f@size{9}\check@mathfonts
\def\maketag@@@#1{\hbox{\m@th\normalsize\normalfont#1}}
\begin{flalign}
\widetilde{c}(\bm{r}_0-\bm{r}')=&\widetilde{h}(\bm{r}_0-\bm{r}')
\nonumber\\
&-\int d\bm{r}"\left\{
\rho^*(\bm{r}")\widetilde{h}(\bm{r}_0-\bm{r}")h(\bm{r}"-\bm{r}')
+\rho^*(\bm{r}")c(\bm{r}_0-\bm{r}")\widetilde{h}(\bm{r}"-\bm{r}')
\right\}+\mathcal{O}[\widetilde{h}^2].
\label{appendix oz app}
\end{flalign}
\endgroup
Substituting Equation~(\ref{appendix oz app}) into Equation~(\ref{partial s}), the~logarithmic contribution $\mathcal{L}(\widetilde{h})$ in Equation~(\ref{partial omega}) becomes
\vspace{-6pt}
\begingroup\makeatletter\def\f@size{9}\check@mathfonts
\def\maketag@@@#1{\hbox{\m@th\normalsize\normalfont#1}}
\begin{flalign}
\mathcal{L}(\widetilde{h})
&=\frac{1}{2}\int\!\!\!\int d\bm{r}_0d\bm{r}'\,\rho^*(\bm{r}_0)\rho^*(\bm{r}')
\widetilde{c}(\bm{r}_0-\bm{r}')\widetilde{h}(\bm{r}_0-\bm{r}')
\nonumber\\
&=\frac{1}{2}\int\!\!\!\int d\bm{r}_0d\bm{r}'\,\rho^*(\bm{r}_0)\rho^*(\bm{r}')
\widetilde{h}^2(\bm{r}_0-\bm{r}')
\nonumber\\
&\qquad
-\frac{1}{2}\int\!\!\!\int\!\!\!\int
 d\bm{r}_0d\bm{r}'d\bm{r}"\,\rho^*(\bm{r}_0)\rho^*(\bm{r}')\rho^*(\bm{r}")
\widetilde{h}(\bm{r}_0-\bm{r}")h(\bm{r}"-\bm{r}')\widetilde{h}(\bm{r}_0-\bm{r}')
\nonumber\\
&\qquad
-\frac{1}{2}\int\!\!\!\int\!\!\!\int
 d\bm{r}_0d\bm{r}'d\bm{r}"\,\rho^*(\bm{r}_0)\rho^*(\bm{r}')\rho^*(\bm{r}")
c(\bm{r}_0-\bm{r}")\widetilde{h}(\bm{r}"-\bm{r}')\widetilde{h}(\bm{r}_0-\bm{r}')
+\mathcal{O}[\widetilde{h}^3].
\end{flalign}
\endgroup
The derivative with respect to $\widetilde{h}(\bm{r})$ gives
\vspace{-6pt}
\begingroup\makeatletter\def\f@size{9}\check@mathfonts
\def\maketag@@@#1{\hbox{\m@th\normalsize\normalfont#1}}
\begin{flalign}
&\frac{\delta\mathcal{L}(\widetilde{h})}{\delta\widetilde{h}(\bm{r})}
\nonumber\\
&\quad\approx
\int\!\!\!\int d\bm{r}_0d\bm{r}'\,
\rho^*(\bm{r}_0)\rho^*(\bm{r}')
\widetilde{h}(\bm{r}_0-\bm{r}')\delta(\bm{r}_0-\bm{r}'-\bm{r})
\nonumber\\
&\quad-\frac{1}{2}\int\!\!\!\int\!\!\!\int
d\bm{r}_0d\bm{r}'d\bm{r}"\,
\rho^*(\bm{r}_0)\rho^*(\bm{r}')\rho^*(\bm{r}")
\nonumber\\
&\qquad\times\left[\left\{
\widetilde{h}(\bm{r}_0-\bm{r}')h(\bm{r}'-\bm{r}")\delta(\bm{r}_0-\bm{r}"-\bm{r})
+\widetilde{h}(\bm{r}_0-\bm{r}")h(\bm{r}"-\bm{r}')\delta(\bm{r}_0-\bm{r}'-\bm{r})
\right\}\right.
\nonumber\\
&\qquad\qquad+\left.\left\{
c(\bm{r}_0-\bm{r}")\widetilde{h}(\bm{r}"-\bm{r}')\delta(\bm{r}_0-\bm{r}'-\bm{r})
+c(\bm{r}"-\bm{r}_0)\widetilde{h}(\bm{r}_0-\bm{r}')\delta(\bm{r}"-\bm{r}'-\bm{r})
\right\}\right],
\label{appendix derivative1}
\end{flalign}
\endgroup
noting that $c(-\bm{r})=c(\bm{r})$.
The approximate relation (\ref{appendix oz app}) can be readily used by rewriting Equation~(\ref{appendix derivative1}) as
\begin{flalign}
\frac{\delta\mathcal{L}(\widetilde{h})}{\delta\widetilde{h}(\bm{r})}
=\frac{1}{2}
\left\{
\int d\bm{r}_0\,\rho^*(\bm{r}_0)\rho^*(\bm{r}_0-\bm{r})E_{\mathrm{oz}}(\widetilde{h})
+\int d\bm{r}'\,\rho^*(\bm{r}'+\bm{r})\rho^*(\bm{r}')E'_{\mathrm{oz}}(\widetilde{h})
\right\},
\label{appendix derivative2}
\end{flalign}
where
\vspace{-6pt}
\begingroup\makeatletter\def\f@size{9}\check@mathfonts
\def\maketag@@@#1{\hbox{\m@th\normalsize\normalfont#1}}
\begin{flalign}
&E_{\mathrm{oz}}(\widetilde{h})\nonumber\\
&=\widetilde{h}(\bm{r})-\int d\bm{r}'\,\rho^*(\bm{r}')
\widetilde{h}(\bm{r}_0-\bm{r}')h(\bm{r}'-\bm{r}_0+\bm{r})
-\int d\bm{r}"\,\rho^*(\bm{r}")
c(\bm{r}_0-\bm{r}")\widetilde{h}(\bm{r}"-\bm{r}_0+\bm{r})
\label{appendix derivative3}
\end{flalign}
\endgroup
and
\vspace{-6pt}
\begingroup\makeatletter\def\f@size{9}\check@mathfonts
\def\maketag@@@#1{\hbox{\m@th\normalsize\normalfont#1}}
\begin{flalign}
&E'_{\mathrm{oz}}(\widetilde{h})\nonumber\\
&=\widetilde{h}(\bm{r})
-\int d\bm{r}"\,\rho^*(\bm{r}")
\widetilde{h}(\bm{r}'+\bm{r}-\bm{r}")h(\bm{r}"-\bm{r}')
-\int d\bm{r}_0\,\rho^*(\bm{r}_0)
c(\bm{r}'+\bm{r}-\bm{r}_0)\widetilde{h}(\bm{r}_0-\bm{r}').
\label{appendix derivative4}
\end{flalign}
\endgroup
It is found from Equation~(\ref{appendix oz app}) that $E_{\mathrm{oz}}(\widetilde{h})=E'_{\mathrm{oz}}(\widetilde{h})\approx\widetilde{c}(\bm{r})$, so that Equation~(\ref{appendix derivative2}) becomes
\begin{flalign}
\frac{\delta\mathcal{L}(\widetilde{h})}{\delta\widetilde{h}(\bm{r})}
=\frac{1}{2}\left\{
\int d\bm{r}_0\,\rho^*(\bm{r}_0)\rho^*(\bm{r}_0-\bm{r})\widetilde{c}(\bm{r})
+\int d\bm{r}'\,\rho^*(\bm{r}'+\bm{r})\rho^*(\bm{r}')\widetilde{c}(\bm{r})
\right\}.
\label{appendix derivative last}
\end{flalign}
Thus, we verify the first term on the rhs of Equation~(\ref{h derivative}) by setting $\bm{r}_0=\bm{r}'+\bm{r}$ in Equation~(\ref{appendix derivative last}).

\section[\appendixname~\thesection]{Derivation of Equation~(\ref{kernel s})}\label{appendix kernel}
The result in Equation~(\ref{kernel s}) is derived as follows:
\vspace{-6pt}
\begingroup\makeatletter\def\f@size{9}\check@mathfonts
\def\maketag@@@#1{\hbox{\m@th\normalsize\normalfont#1}}
\begin{flalign}
\mathcal{M}(\bm{k},\bm{q})
&=\frac{1}{S^2(\bm{q})\,S^2(\bm{k}-\bm{q})}
-\frac{1}{S^2(\bm{q})}
-\frac{1}{S^2(\bm{k}-\bm{q})}
\nonumber\\
&=\left\{\frac{1}{S^2(\bm{q})}-1\right\}\left\{\frac{1}{S^2(\bm{k}-\bm{q})}-1\right\}-1
\nonumber\\
&=\left\{\overline{\rho}^2c^2_*(\bm{q})-2\overline{\rho}c_*(\bm{q})\right\}\left\{\overline{\rho}^2c^2_*(\bm{k}-\bm{q})-2\overline{\rho}c_*(\bm{k}-\bm{q})\right\}-1
\nonumber\\
&=\left\{\overline{\rho}^2c_*(\bm{q})c_*(\bm{k}-\bm{q})\right\}^2+2\overline{\rho}^2c_*(\bm{q})c_*(\bm{k}-\bm{q})\left\{2-\overline{\rho}c_*(\bm{q})-\overline{\rho}c_*(\bm{k}-\bm{q})\right\}-1
\nonumber\\
&=\left\{\overline{\rho}^2c_*(\bm{q})c_*(\bm{k}-\bm{q})\right\}^2+2\overline{\rho}^2c_*(\bm{q})c_*(\bm{k}-\bm{q})\left\{\frac{1}{S(\bm{q})}+\frac{1}{S(\bm{k}-\bm{q})}\right\}-1,
\end{flalign}
\endgroup
where use has been made of the relation, $S(\bm{q})=\left\{1-\overline{\rho}c_*(\bm{q})\right\}^{-1}$, in~the last~line.

\end{document}